\documentclass[review,numbers,sort&compress]{elsarticle}

\usepackage{lineno,hyperref}
\modulolinenumbers[5]

\usepackage{geometry}
\journal{Elsevier}

\hypersetup{hidelinks}
\graphicspath{{pic/}}
\date{}
\usepackage{booktabs}
\usepackage{siunitx}
\usepackage{amsmath}
\usepackage{subfigure}
\usepackage{float}
\usepackage[capitalize]{cleveref}
\usepackage{xcolor}
\usepackage{multirow}
\usepackage{amssymb}
\usepackage[linesnumbered,algoruled,boxed,lined]{algorithm2e}
\usepackage{enumitem}
\setlist[description]{font=\normalfont}
 
\usepackage{caption}
\newcommand{\blueblack}{\color{black}}

\begin{document}

\begin{frontmatter}

\title{A validated fluid-structure interaction simulation model for vortex-induced vibration of a flexible pipe in steady flow}

\author[mymainaddress]{Xuepeng Fu}
%\ead{xuepeng.fu@nrel.gov; isfuxp@sjtu.edu.cn}
%\fntext[myfootnote]{Current Address: National Renewable Energy Laboratory, Golden, Colorado, 80401, USA.}

\author[mymainaddress]{Shixiao Fu\corref{mycorrespondingauthor}}
\cortext[mycorrespondingauthor]{Corresponding author}
\ead{shixiao.fu@sjtu.edu.cn}

\author[mymainaddress]{Zhibo Niu}

\author[mymainaddress]{Bing Zhao}
\author[mymainaddress]{Jiawei Shen}
\author[mymainaddress]{Pengqian Deng}

\address[mymainaddress]{State Key Laboratory of Ocean Engineering, Shanghai Jiao Tong University, Shanghai, 200240, China}

%\address[mythirdaddress]{National Renewable Energy Laboratory, Golden, CO, 80401, USA}

\begin{abstract}
{\blueblack We propose a validated fluid--structure interaction simulation framework based on the strip method for the vortex-induced vibration of a flexible pipe. The numerical results are compared with the experimental data from three steady flow conditions: uniform, linearly sheared, and bidirectionally sheared flow. The Reynolds number ranges from $10^4$ to $10^5$. The flow field is simulated based on open-source software OpenFOAM. The solid field is modeled based on the finite element method of the Euler--Bernoulli beam, and fluid--structure coupling is implemented via a weak coupling algorithm developed in MATLAB. The vortex-induced vibration response is assessed in terms of amplitude and frequency, along with the differences in strain. Additionally, wavelet analysis and traveling wave phenomena are investigated. This study presents the first numerical simulation of flexible pipe VIV under bidirectionally sheared flow, validated against experimental data. Compared to uniform and linearly sheared flow, the bidirectionally sheared flow condition leads to more pronounced traveling wave behavior and stronger multi-frequency responses, especially in the in-line direction. The simulation results are directly compared with measured strain data, showing agreement across different flow conditions. The numerical simulation codes and experimental data in this manuscript are openly available, providing a foundation for more complex vortex-induced vibration simulations in the future.}
\end{abstract}

\begin{keyword}
vortex-induced vibration\sep flexible pipe\sep numerical simulation \sep steady flow
\end{keyword}

\end{frontmatter}
%\linenumbers

{\blueblack
\section*{Nomenclature}
\begin{description}[align=left,labelwidth=5em,labelsep=3em]
  \item[VIV] Vortex-induced vibration
  \item[IL] In-line
  \item[CF] Cross flow
  \item[CFD] Computational fluid dynamics
  \item[FSI] Fluid-structure interaction
  \item[URANS] Unsteady Reynolds-averaged Navier-Stokes
  \item[FEM] Finite element method
  \item[PSD] Power spectral density
  \item[$Re$] Reynolds number
  \item[$u_i$] Fluid velocity component ($i = 1,2,3$)
  \item[$p$] Pressure
  \item[$\rho$] Density of fluid
  \item[$\nu$] Kinematic viscosity of the fluid
  \item[$\mu$] Dynamic viscosity of the fluid 
  \item[$\boldsymbol{F}_f$] Fluid force acting on the structure
  \item[$\boldsymbol{I}$] Identity tensor
  \item[$\boldsymbol{R}_{dev}$] Deviatoric Reynolds stress tensor
  \item[$\boldsymbol{S}$] Face area vector
  \item[$\boldsymbol{M}$] System mass matrix
  \item[$\boldsymbol{C}$] System damping matrix
  \item[$\boldsymbol{K}$] System stiffness matrix
  \item[$\boldsymbol{\xi}$] Structural displacement matrix
  \item[$\mathcal{N}(\cdot)$] Fluid domain governing equations
  \item[$\mathcal{S}(\cdot)$] Structural domain governing equations
  \item[$U$] Flow velocity in uniform flow cases
  \item[$U_{max}$] Maximum flow velocity in linearly and bidirectionally sheared flow cases
  \item[$L$] Flexible pipe length
  \item[$D$] Flexible pipe diameter
  \item[$\bar{m}$] Unit length mass of flexible pipe in air 
  \item[$EI$] Bending stiffness
  \item[$EA$] Tensile stiffness
  \item[$\zeta$] Damping ratio
  \item[$Z/L$] Nondimensional spanwise location
  \item[$A_{ini}/D$] Nondimensional IL initial displacement induced by mean drag
  \item[$A_{IL}/D$] Nondimensional RMS IL VIV displacement
  \item[$A_{CF}/D$] Nondimensional RMS CF VIV displacement
  \item[$\varepsilon_{IL}$] RMS IL strain
  \item[$\varepsilon_{CF}$] RMS CF strain
  \item[$\mathcal{WT}$] Wavelet transformation 
  \item[$f$] Frequency component
  \item[$\mathcal{F}(f)$] General amplitude-frequency spectrum
  \item[$\hat{F}(f)$] Power spectral density
\end{description}
}

\section{Introduction}
Vortex-induced vibration (VIV) is a critical phenomenon in offshore engineering, especially for flexible structures such as risers, mooring lines, and pipelines \cite{huera2024vortex}. VIV occurs when fluid flows past these structures, generating a vortex street that induces mechanical oscillations in both the in-line (IL) and cross flow (CF) directions. If not properly considered in the design and operation of these structures, VIV can lead to fatigue damage and structural failure. 

The prediction of vortex-induced vibration is a crucial aspect of VIV research, with researchers in both academia and industry having explored this field for nearly half a century. The VIV studies began with bluff cylinders \cite{govardhan2000modes, dahl2006two}. However, in engineering fields, especially ocean engineering, flexible pipeline structures are widely used. The multi-frequency response and traveling wave characteristics of flexible structures make the VIV response more complex than that of bluff structures. Flexible pipe VIV is a typical fluid-structure interaction (FSI) phenomenon. Therefore, response prediction consists of two parts: the flexible pipe structure model and the flow field model. For the flexible pipe model, Euler-Bernoulli beams with tension and catenary models have been widely validated as effective. With respect to the flow field model, many different low and high fidelity models have been proposed, such as the simple Morison equation and wake oscillator models \cite{qu2020single,soares2021modelling}, as well as those based on forced motion hydrodynamic datasets \cite{vandiver2017does,lu2018modal,thorsen2016time} and computational fluid dynamics (CFD). The flow field model plays a crucial role in simulating the VIV phenomenon. Moreover, with the recent development of the submerged floating tunnel, research on coupling schemes, such as weak and strong coupling schemes, has emerged in VIV community.

All VIV response prediction methods for flexible risers must be validated through experiments. However, flexible pipes typically have diameters on the centimeter scale, and their lengths can extend from tens to hundreds of meters \cite{trim2005experimental,tognarelli2004viv,chaplin2005laboratory,vandiver2005high,resvanis2023report}. Previous experimental studies focused primarily on the structural response, with a limited focus on the fluid domain. While some studies have attempted to derive hydrodynamic forces on the basis of structural responses \citep{song2016investigation, wu2010estimation}, there remains a lack of knowledge regarding fluid domain data. Given the important role of viscosity, potential flow theory is insufficient for this purpose. In essence, only CFD methods based on the Navier-Stokes equations can provide meaningful insights into the fluid domain.

{\blueblack For the vortex-induced vibration of flexible risers, high-fidelity three-dimensional direct numerical simulation simulations are typically limited to Reynolds number ($Re$) on the order of hundreds \citep{bourguet2011vortex,fan2019mapping,wang2021large}. At these flow speeds, the noise associated with measurement equipment is high, which makes DNS simulations suitable for only purely theoretical research. Few commercial software packages such as ANSYS currently have the capability to simulate vortex-induced vibrations of flexible pipes \citep{chen2008numerical}. The strip method provides a feasible approach for studying high Reynolds numbers. Few famous CFD strip methods for flexible pipes \citep{han2022surface,wang2022numerical}, from Norsk Hydro \citep{HerfjordAss} and VIVIC \citep{willden2004multi} to Nektar++ \citep{bao2016generalized}, have been investigated. Many studied have been conducted with flexible pipe VIV simulation \citep{wang2016numerical,huang2011numerical}. However, these methods have two major drawbacks: 1) Flexible riser experiments are extremely costly, making widespread validation of any proposed model difficult. Additionally, the most famous published blind validation was conducted decades ago \citep{chaplin2005blind}. 2) The software for these methods is closed-source, which presents a significant barrier to conducting CFD simulations of flexible risers \citep{deng2021numerical}. }

In this paper, we propose an open-source fluid-structure interaction simulation model for the VIV of a flexible pipe based on the strip method. Validation studies are conducted using three steady flow conditions from previous experimental research. We aim to initiate the open-source development of CFD simulations for flexible pipe VIV. All the experimental data and codes used in this study are hosted on GitHub.

\section{Numerical methods}
\subsection{Fluid domain}

The governing equation of the fluid domain is the incompressible Navier-Stokes (N-S) equation, which can be written as:
\begin{align}
& \frac{\partial u_i}{\partial x_i}=0, \\
& \frac{\partial u_i}{\partial t}+\frac{\partial u_i u_j}{\partial x_j}=-\frac{1}{\rho} \frac{\partial p}{\partial x_i}+\nu \frac{\partial^2 u_i}{\partial x_j \partial x_j},
\end{align}
where $u_i$ represents the fluid velocity, $x_i$ represents the coordinates, with the subscript $i = 1,2,3$ denoting the $x$, $y$, and $z$ components, $p$ represents the pressure, $\rho$ represents the fluid density, and $\nu$ represents the kinematic viscosity of the fluid.

\begin{figure}[ht!]
    \centering
    \includegraphics[width=.8\textwidth]{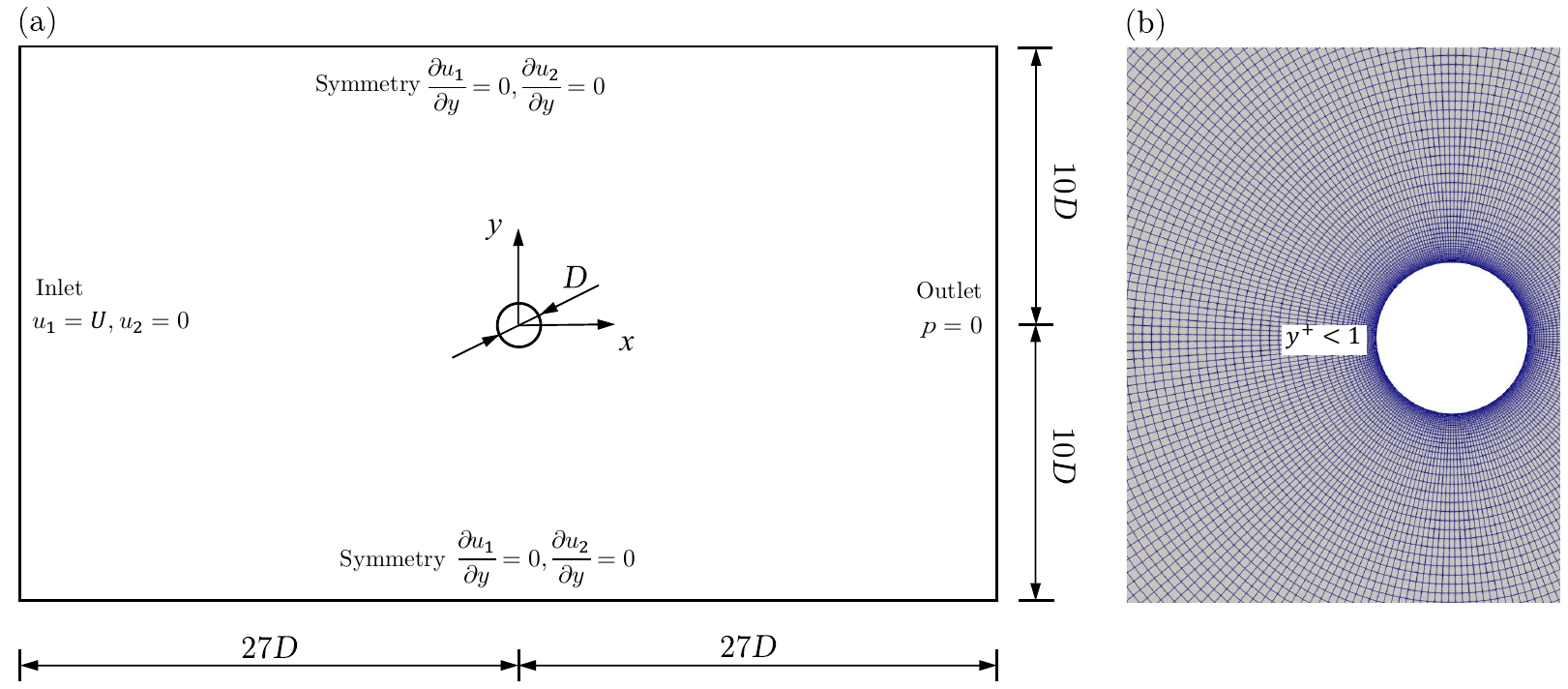}
    \caption{Sketch of strip initial mesh: (a) Fluid domain size and boundary condition set-up; (b) near wall mesh with $y^+<1$. }
  \label{meshpic}
\end{figure}

The above equations serve as the control equations for the fluid domain. Numerical solutions of these equations provide the velocity and pressure distributions within the flow field. However, directly solving these equations requires substantial computational resources, as the grid density is proportional to $Re^{2.6} $\citep{pope2000turbulent}. Therefore, the Reynolds-averaged Navier-Stokes (RANS) equations are employed to simplify the solution of the N-S equations. Applying time averaging to the N-S equations yields the following RANS equations:
\begin{align} 
& \frac{\partial \bar{u}_i}{\partial x_i}=0, \label{ranscon}\\
& \frac{\partial \bar{u}_i}{\partial t}+\frac{\partial \bar{u}_i \bar{u}_j}{\partial x_j}=-\frac{1}{\rho} \frac{\partial \bar{p}}{\partial x_i}+\nu \frac{\partial^2 \bar{u}_i}{\partial x_j \partial x_j}-\frac{\partial \overline{u_i^{\prime} u_j^{\prime}}}{\partial x_j}, \label{ransmom}
\end{align}
where $ \bar{(\cdot)} $ denotes the averaging operator and $ \overline{u_i^{\prime} u_j^{\prime}} $ represents the Reynolds stress, which signifies energy transfer due to turbulent fluctuations. In the RANS equations, the Reynolds stress requires closure modeling, and the SST $ k-\omega $ turbulence model \citep{menter2003ten} is applied for simulation. {\blueblack The open-source partial differential equation solver package based on the finite volume method, OpenFOAM-8 \citep{jasak2007openfoam}, is applied in the present study for simulating the unsteady Reynolds-averaged Navier-Stokes (URANS) equations to obtain the fluid force distribution:
\begin{equation}
\boldsymbol{F}_f=\oint\left(p\boldsymbol{I}+\mu\boldsymbol{R}_{dev}\right)\cdot\mathrm{d}\boldsymbol{S},
\end{equation}
where $p$ is the pressure, $\boldsymbol{R}_{dev}$ is the deviatoric Reynolds stress tensor\citep{fu2024data}, $\mu$ is the dynamic viscosity of fluid, and $\boldsymbol{S}$ is the face area vector. All temporal and spatial schemes are second-order. For further details on the numerical schemes, refer to previous studies \citep{fu2022frequency,fu2023numerical,fu2024data}. \cref{meshpic} illustrates the computational domain and boundary conditions adopted for the fluid simulation. The domain is designed to be sufficiently large to avoid boundary effects, with a total length of $54D$ and height of $20D$, ensuring that the flow around the structure develops fully before reaching the outlet. A uniform inflow condition is applied at the inlet, where $u_1 = U$ and $u_2 = 0$, and a zero-pressure condition ($p = 0$) is imposed at the outlet. Symmetry boundary conditions are applied on the top and bottom walls to emulate an unbounded domain in the vertical direction, where the normal gradients of velocity components vanish, i.e., $\partial u_1 / \partial y = \partial u_2 / \partial y = 0$. The computational mesh near the wall is refined to ensure that the non-dimensional wall distance satisfies $y^+ < 1$, allowing accurate resolution of the viscous sublayer. Furthermore, the time step is selected such that the Courant number remains below 1 throughout the simulation to ensure numerical stability and accuracy.
}

\subsection{Solid domain}
The tensioned flexible pipe undergoing vortex-induced vibration is typically simulated with the Euler-Bernoulli beam model \citep{larsen1995generalized,vandiver2005high}. In this study, the finite element method (FEM) is used to solve for the structural response. The governing equation for the structure is as follows:
\begin{equation}
    \boldsymbol{M}\ddot{\boldsymbol{\xi}}+\boldsymbol{C}\dot{\boldsymbol{\xi}}+\boldsymbol{K}\boldsymbol{\xi} = \boldsymbol{F}_{f},
\end{equation}
where  $\boldsymbol{M}$, $\boldsymbol{C}$, and $\boldsymbol{K} $ are the structural system mass matrix, damping matrix, and stiffness matrix of the structural system, respectively, $ \boldsymbol{\xi} $ is the displacement matrix, and $ \boldsymbol{F}_{f} $ is the fluid force acting on the structure. More details about the FEM model are provided in \ref{appenfem}.

The Newmark-$ \beta $ method is used to obtain get the time-domain response, with constants $ \beta = 0.25 $ and $ \gamma = 0.5 $. Based on the defined simulation time step $ \Delta t $, the following constants are computed:
\begin{equation}
\begin{aligned}
&a_0 =\frac{1}{\beta\Delta t^{2}},\quad a_{1}=\frac{\gamma}{\beta\Delta t},\quad a_{2}=\frac{1}{\beta\Delta t},\quad a_{3}=\frac{1}{2\beta}-1,\quad a_{4}=\frac{\gamma}{\beta}-1,  \\
&a_5 =\frac{\Delta t}{2} \left(\frac{\gamma}{\beta}-2\right),\quad a_{6}=\Delta t(1-\gamma),\quad a_{7}=\gamma\Delta t,
\end{aligned}    
\end{equation}
and the effective stiffness matrix is constructed as:
\begin{equation}
\boldsymbol{K}^*=\boldsymbol{K}+a_0\boldsymbol{M}+a_1\boldsymbol{C},
\end{equation}
along with the load vector:
\begin{equation}
\boldsymbol{F}_{t+\Delta t }^*=\boldsymbol{F}_{t+\Delta t}+\boldsymbol{M}\left(a_0\boldsymbol{\xi}_t+a_2\dot{\boldsymbol{\xi}}_t+a_3\ddot{\boldsymbol{\xi}}_t\right)+\boldsymbol{C}\left(a_1\boldsymbol{\xi}_t+a_4\dot{\boldsymbol{\xi}}_t+a_5\ddot{\boldsymbol{\xi}}_t\right).   
\end{equation}
The displacement matrix of the riser at the next time step is obtained via:
\begin{equation}
\begin{aligned}
& \boldsymbol{\xi}_{t+\Delta t}=\boldsymbol{K}^{*-1}\boldsymbol{F}_{t+\Delta t }^*,\\
&\dot{\boldsymbol{\xi}}_{t+\Delta t} =\dot{\boldsymbol{\xi}}_t+a_6\ddot{\boldsymbol{\xi}}_t+a_7\ddot{\boldsymbol{\xi}}_{t+\Delta t}, \\
&\ddot{\boldsymbol{\xi}}_{t+\Delta t} =a_0(\boldsymbol{\xi}_{t+\Delta t}-\boldsymbol{\xi}_t)-a_2\dot{\boldsymbol{\xi}}_t-a_3\ddot{\boldsymbol{\xi}}_t.  
\end{aligned}
\end{equation}

\subsection{Fluid-structural interaction coupling algorithm}
VIV is a typical fluid-structure interaction phenomenon. Solving fluid-structure interaction problems requires exchanging fluid force information $\boldsymbol{F}_{f}$ and structural displacement information $\boldsymbol{\xi}_s$ between the solid and fluid domains. By considering the governing equations in both domains and incorporating boundary conditions at the fluid-structure interface, the fluid-structure coupling governing equations can be expressed as:
\begin{equation}
\left\{
\begin{matrix}
\begin{aligned}
\mathcal{N}(\boldsymbol{u}_f,p_f)=0&\qquad\text{in}\quad\Omega_f,\\ 
\mathcal{S}(\boldsymbol{\xi}_s)-\boldsymbol{F}(\boldsymbol{u}_f,p_f)=0&\qquad\text{in}\quad\Omega_s,\\
\boldsymbol{u}_f=\boldsymbol{u}_s&\qquad\text{on}\quad\Gamma_{f,s},\\ 
\boldsymbol{\sigma}_f=\boldsymbol{\sigma}_s&\qquad\text{on}\quad\Gamma_{f,s},\\ 
\end{aligned}
\end{matrix}\right.    
\end{equation}
where $\mathcal{N}(\cdot)$ represents the fluid domain governing equations, $\mathcal{S}(\cdot)$ represents the structural domain governing equations, $\Omega_f$ and $\Omega_s$ are the fluid and solid domains, respectively, and $\Gamma_{f,s}$ denotes the fluid-structure interface. Fluid-structure coupling solutions typically involve weak or strong coupling approaches. In a weak coupling (or partitioned) approach, the equations for the fluid and solid domains are solved separately, with a single data exchange step at each coupling step. In a strong coupling, the equations in both domains are solved simultaneously, ensuring the satisfaction of boundary conditions, or separately with multiple data exchange steps.

In the present study, the weak FSI coupling algorithm, as shown in Algorithm \ref{alg:algorithmweak}, solves the fluid and solid equations independently, with a single data exchange at each coupling step. The weak coupling algorithm is applied in this manuscript. In contrast, a strong coupling algorithm, which performs multiple iterations within a single time step, is also considered. The differences between the two methods are discussed in \ref{weakstrong}.

\begin{algorithm}[ht!]
	\caption{Weak coupling algorithm of FSI simulation}
	\label{alg:algorithmweak}
	\KwIn{Solid field parameters; Flow field parameters.}
	\KwOut{Structural response $\boldsymbol{\xi}_s$ and flow velocity and pressure field $\boldsymbol{u}_f,p_f$.}  
	$t \leftarrow 0$
	
	\While{$t<t_{end}$}{
	$\boldsymbol{F}_{f}$ $\leftarrow$ \textnormal{solve} $\mathcal{N}(\boldsymbol{u}_f,p_f)=0$
	
	 $\boldsymbol{\xi}_s$ $\leftarrow$ \textnormal{solve} $\mathcal{S}(\boldsymbol{u}_s)-\boldsymbol{F}_{f}=0 $
  
    \textnormal{Update mesh with} $\boldsymbol{\xi}_s$
    
	$t  \leftarrow t+\Delta t$
	}
\end{algorithm}

\section{Benchmark VIV experiments}
The benchmark experiments referenced in this manuscript have already been conducted and published by our research group. We only introduced some critical information for the CFD simulation in this manuscript; more details are provided in the references.

\subsection{Uniform flow cases}\label{uni}
The uniform flow VIV experiment was conducted in the towing tank of the Shanghai Ship and Shipping Research Institute. The flexible riser model was constructed from a brass pipe, coated with a heat-shrink tube to protect the embedded Fiber Bragg Grating (FBG) sensors. This experiment, conducted over a decade ago, utilized a brass riser model with high bending stiffness $EI$, which required high flow velocities to excite higher-mode VIV. As a result, the experimental conditions are not well suited for direct comparison with CFD simulations. For our simulations, two cases were selected from this experiment. Further details about the flexible pipe experiments can be found in related studies \citep{ren2022hydrodynamic}. A pretension force $\SI{3000}{N}$ was applied to the pipe model through the tensioner.

\begin{figure}[ht!]
    \centering
    \includegraphics[width=.65\textwidth]{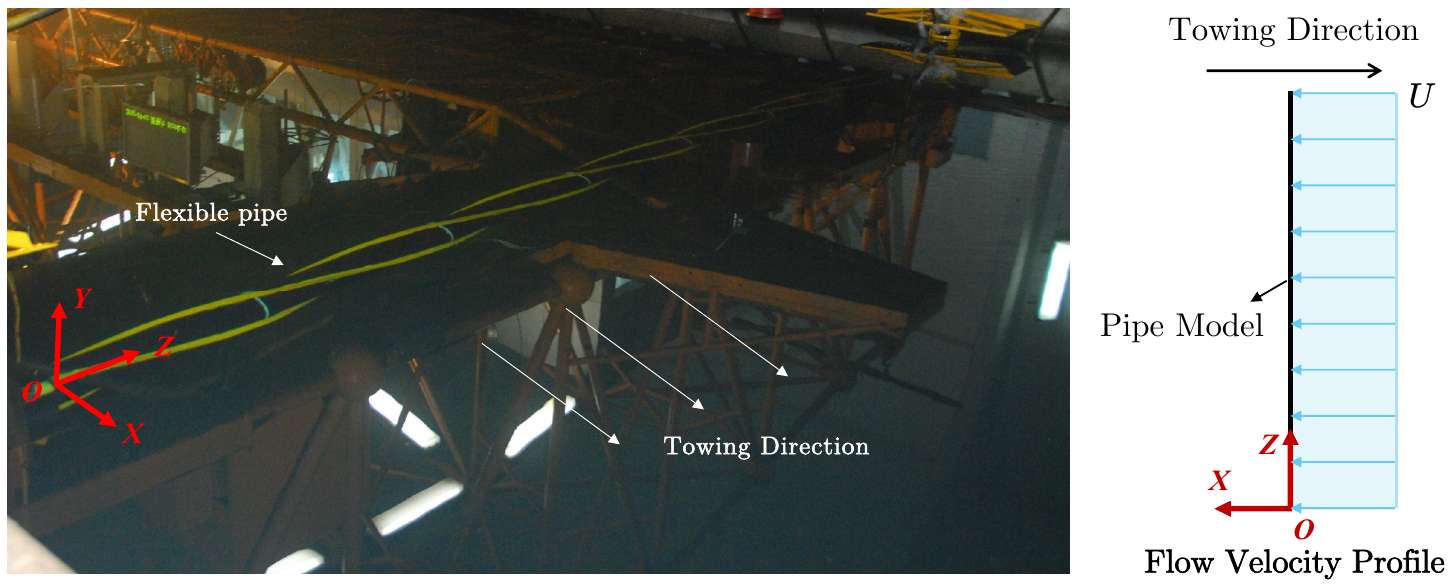}
    \caption{Experimental setup of the flexible pipe VIV test in uniform flow. The uniform flow is generated by passive towing.}
  \label{uniformflowtest}
\end{figure}

The primary physical properties of the test pipe model in the uniform flow case are listed in \cref{tabphyuni}.  

\begin{table}[h!]
	\centering
	\caption{Physical properties of the test pipe model in uniform flow case}
	\label{tabphyuni}
	\begin{tabular}{@{}ll@{}}
		\toprule
		Parameter                                                                 & Value of test model \\ \midrule
		Pipe model length $L$ (m)                                               & 7.90  \\
		Outer diameter $D$ (mm)                                                & 31.00 \\
		Mass in air $\bar{m}$ (kg/m)                          & 1.77  \\
		Bending stiffness $EI$ ($\SI{}{Nm^2}$) & $1.47\times 10^3$  \\
		Tensile stiffness $EA$ (N)                                              & $1.45\times 10^3$ \\
		Damping ratio $\zeta$ (\%)                              & 0.3  \\ \bottomrule
	\end{tabular}
\end{table}

\subsection{Linearly sheared flow cases}\label{linear}
\begin{figure}[ht!]
    \centering
    \includegraphics[width=.65\textwidth]{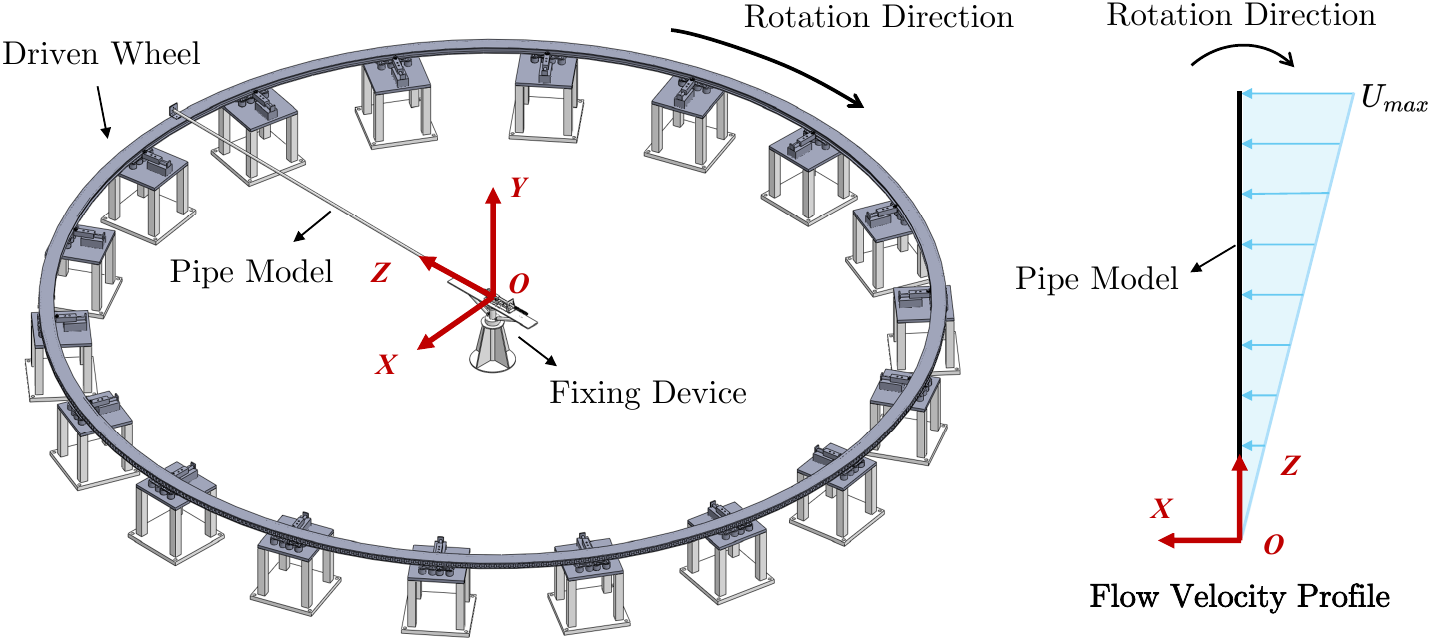}
    \caption{Experimental setup of the flexible pipe VIV test in linearly sheared flow. The linearly sheared flow is generated by the driven wheel rotation. This figure is modified based on the figure in the previous paper \citep{fu2024vortex}. }
  \label{lishearflowtest}
\end{figure}

The model test was conducted in the ocean basin of the State Key Laboratory of Ocean Engineering at Shanghai Jiao Tong University. The experimental setup, consisting of the rotating rig and the tested pipe model, was installed on the false bottom of the basin, as illustrated in \cref{lishearflowtest}. The flow field was generated by rotating the apparatus via a timing belt driven by a servo motor, and the driven wheel moved at the same given linear velocity via the servo motor during the experiment. This novel VIV experimental apparatus has undergone credibility validation through noise signal analysis, repetitive experiments, and water depth independence tests, as stated in previous studies \citep{fu2022experimental,fu2022study}. Unlike the VIV experiment for bidirectional shear flow discussed later, in this case, a device was positioned at the center of the driven wheel to secure the wheel. The pipe model was tensioned between the edge of the apparatus at the driven wheel and the central fixed device, which was equipped with clamps, U-joints, and a force sensor. The distance from the false bottom to the experimental pipe was $26 D$. A pretension force of $\SI{550}{N}$ was applied to the pipe model.

The primary physical properties of the test pipe model in the linearly sheared flow case are listed in \cref{tabphylishear}.  

\begin{table}[h!]
	\centering
	\caption{Physical properties of the test pipe model in linearly sheared flow case}
	\label{tabphylishear}
	\begin{tabular}{@{}ll@{}}
		\toprule
		Parameter                                                                 & Value of test model \\ \midrule
		Pipe model length $L$ (m)                                               & 3.88  \\
		Outer diameter $D$ (mm)                                                & 28.41 \\
		Mass in air $\bar{m}$ (kg/m)                          & 1.24  \\
		Bending stiffness $EI$ ($\SI{}{Nm^2}$) & $58.6$  \\
		Tensile stiffness $EA$ (N)                                              & $9\times 10^5$ \\
		Damping ratio $\zeta$ (\%)                              & 2.58  \\ \bottomrule
	\end{tabular}
\end{table}

\subsection{Bidirectionally sheared flow cases}\label{bilinear}

\begin{figure}[ht!]
    \centering
    \includegraphics[width=.65\textwidth]{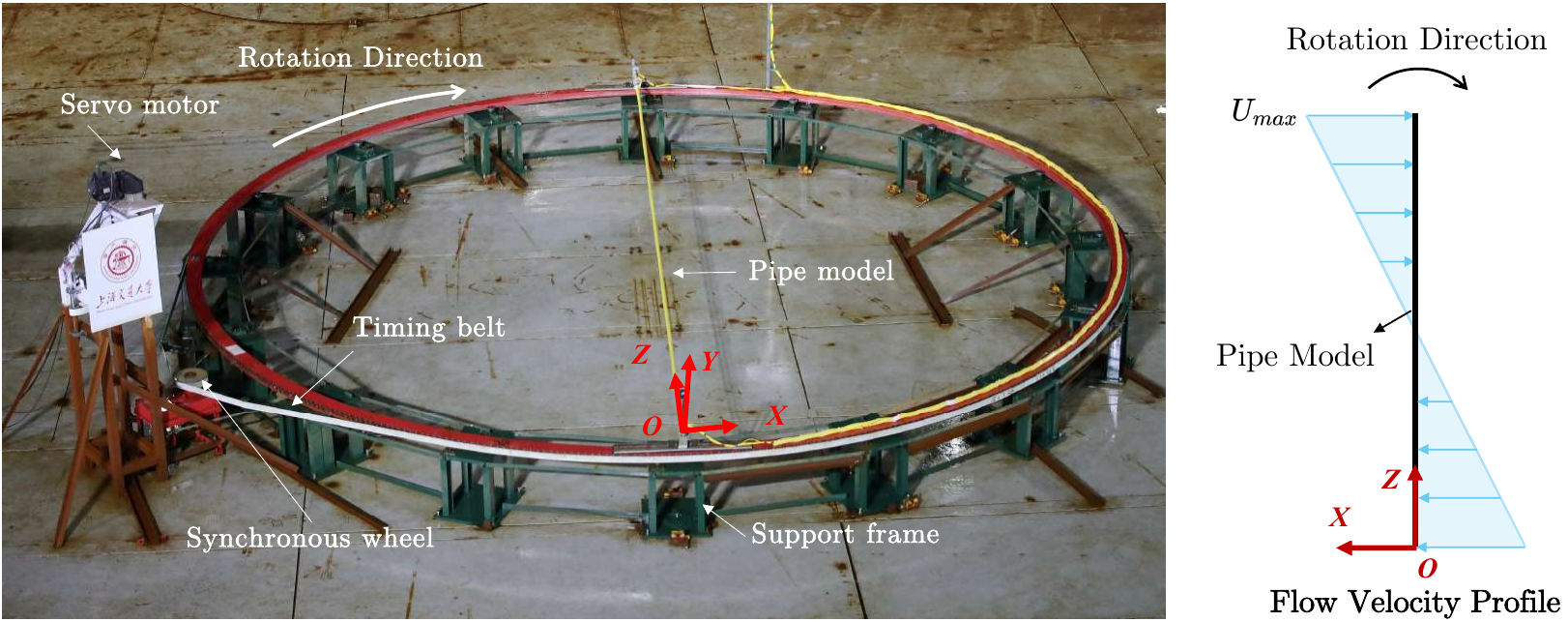}
    \caption{Experimental setup of the flexible pipe VIV test in bidirectionally sheared flow. The bidirectionally sheared flow was generated by the rotation of the driven wheel. This figure is modified from a figure in the previous paper \citep{fu2022experimental}. }
  \label{bishearflowtest}
\end{figure}

The model test was performed in an ocean basin at Shanghai Jiao Tong University. The experimental apparatus was similar to that used in previous linearly sheared flow tests. The test pipe model was installed on two sides through the diameter of the driven gear, and it consisted of clamps, U-joints and a force sensor, as shown in \cref{bishearflowtest}. The force sensor was connected to a tensioner fixed on the driven gear. A pretension force $\SI{980}{N}$ was applied to the pipe model through the tensioner. 

The primary physical properties of the test pipe model in the bidirectionally sheared flow case are listed in \cref{tabphybi}.

\begin{table}[h!]
	\centering
	\caption{Physical properties of the test pipe model in bidirectionally sheared flow case}
	\label{tabphybi}
	\begin{tabular}{@{}ll@{}}
		\toprule
		Parameter                                                                 & Value of test model \\ \midrule
		Pipe model length $L$ (m)                                               & 7.64  \\
		Outer diameter $D$ (mm)                                                & 28.41 \\
		Mass in air $\bar{m}$ (kg/m)                          & 1.24  \\
		Bending stiffness $EI$ ($\SI{}{Nm^2}$) & 58.6  \\
		Tensile stiffness $EA$ (N)                                              & 9.4E5 \\
		Damping ratio $\zeta$ (\%)                              & 2.58  \\ \bottomrule
	\end{tabular}
\end{table}

\section{Results and discussion}
In this manuscript, we introduce the vibration response results and compare three kinds of flow cases; then, a hydrodynamic analysis of two kinds of flow cases, uniform flow and linearly sheared flow, is illustrated. The hydrodynamic analysis of bidirectionally sheared flow cases will be introduced later because of its complexity.
\subsection{Uniform flow cases}
\begin{figure}[ht!]
    \centering
    \includegraphics[width=.85\textwidth]{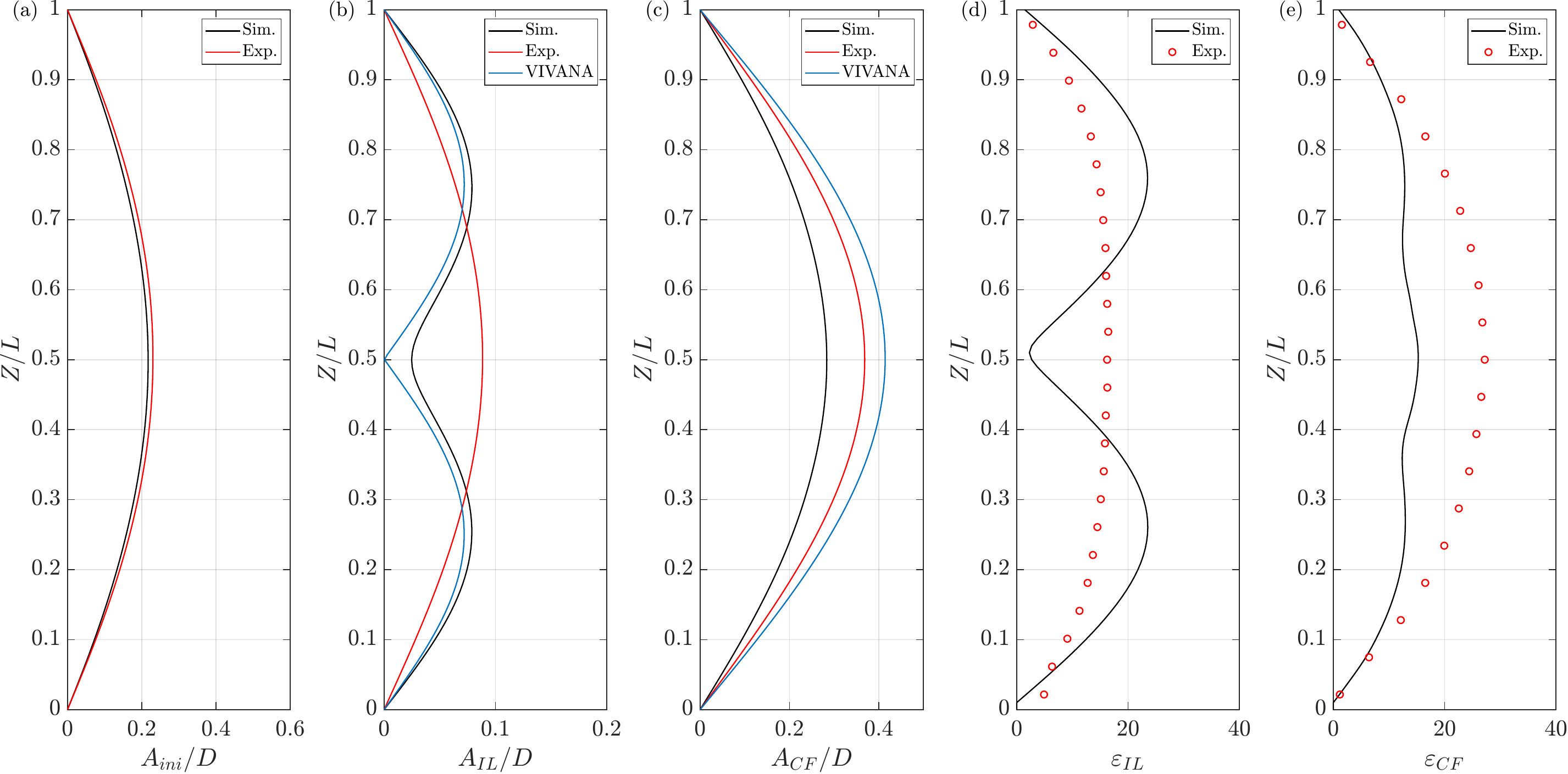}
    \caption{Spanwise distribution of VIV response of uniform flow case at $U=\SI{0.40}{m/s}$: (a) initial displacement in the IL direction; (b) VIV RMS displacement in the IL direction; (c) VIV RMS displacement in the CF direction; (d) VIV RMS strain in the IL direction; (e) VIV RMS strain in the CF direction. Black line: simulation result; red line and circular symbol: experimental result; blue line: VIVANA result.}
  \label{uniformres040}
\end{figure}

For uniform flow cases, we introduce two cases: flow velocities of $U = \SI{0.40}{m/s}$ and $\SI{1.60}{m/s}$. \cref{uniformres040} represents the distribution of the VIV response in the $U=\SI{0.40}{m/s}$ case with five subfigures. The subfigure~(a) represents the initial displacement of the flexible pipe due to the mean drag force; subfigures~(b) and (c) represent the VIV responses in the CF and IL directions, respectively; and subfigures~(d) and (e) represent the strain data, respectively. The black line represents the simulation result; the red line represents the experimental result; the blue line represents the VIVANA prediction result; and the red dots represent the strain data measured in the experiment. All the data points are calculated based on the last $20\%$ data, and the VIVANA is a frequency domain model with no time-domain data.

VIVANA \citep{larsen2012frequency} is applied in this study to compare with the prediction results, as it is also a strip theory based method, and the nodal forces are derived from a forced motion VIV database for a bluff cylinder. As a widely used engineering VIV prediction platform, VIVANA is employed here for the preliminary validation of the CFD numerical results. However, the vortex-induced vibration of a flexible pipe is a highly complex FSI phenomenon, and no prediction software can perfectly replicate the experimental results. Fatigue prediction, which is critical for engineering applications, depends on the displacement, frequency, and traveling wave phenomena. We introduce these results in this section.

\begin{figure}[ht!]
    \centering
    \includegraphics[width=.8\textwidth]{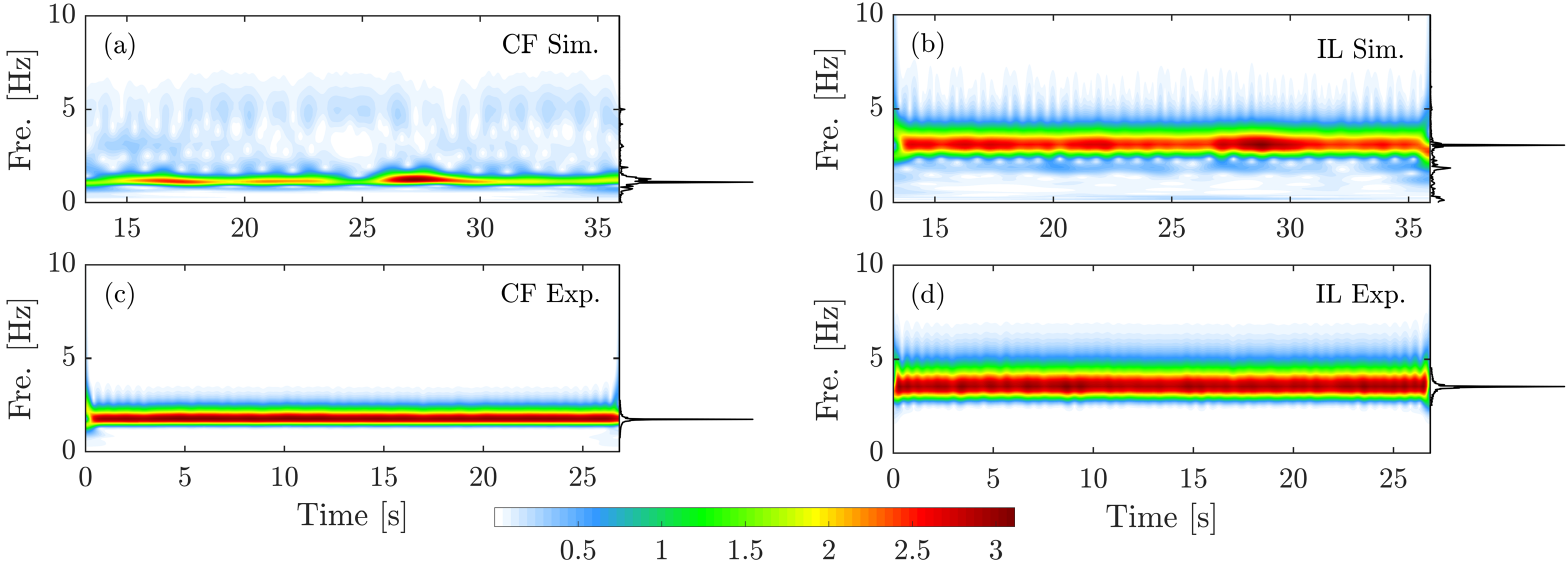}
    \caption{Response frequencies at $Z = 0.2L$ of uniform flow $U=\SI{0.40}{m/s}$ case: (a) time-varying response frequency obtained via a wavelet analysis of the numerical result in the CF direction; (b) time-varying response frequency of the numerical result in the IL direction; (c) time-varying response frequency of the experimental result in the CF direction; (d) time-varying response frequency of the experimental result in the IL direction. Black line on the right represents the general frequency spectrum. White to red: zero to maximum value.}
  \label{uniform040timefre}
\end{figure}

In most flexible pipe VIV experiments, strain signals are measured initially, followed by modal analysis to obtain the VIV displacement. In this study, numerical displacement is converted to strain using the following relationship:
\begin{equation}
\varepsilon_{CF} = \dfrac{\partial^2 y}{\partial z^2}R,
\end{equation}
where $R$ represents the radius where the strain gauge is positioned. $\varepsilon_{IL}$ can be obtained via the same method. Experimental strain data are directly obtained during the tests. For the uniform flow case, a bandpass filter is applied during postprocessing, whereas in the linearly sheared and bidirectionally sheared flow cases, unfiltered raw data are used.

\begin{figure}[ht!]
    \centering
    \includegraphics[width=.85\textwidth]{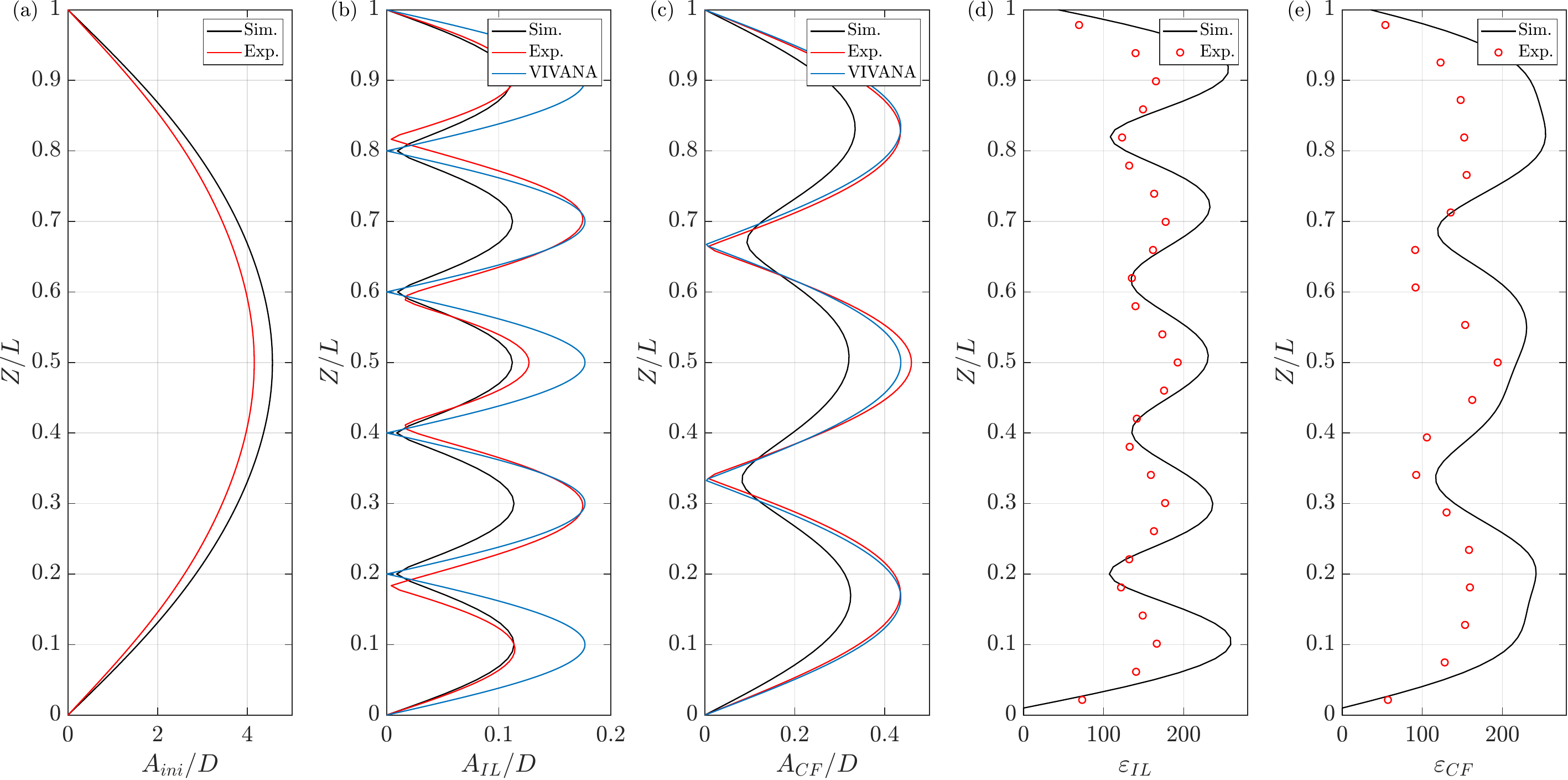}
    \caption{Spanwise distribution of VIV response of uniform flow case at $U=\SI{1.60}{m/s}$: (a) initial displacement in the IL direction; (b) VIV RMS displacement in the IL direction; (c) VIV RMS displacement in the CF direction; (d) VIV RMS strain in the IL direction; (e) VIV RMS strain in the CF direction. Black line: simulation result; red line and circular symbol: experimental result; blue line: VIVANA prediction result.}
  \label{uniformres160}
\end{figure}

However, it should be noted that the modal analysis approach used in this case is still being researched
and developed \citep{chaplin2005laboratory, ren2020drag, huarte2006multi}. There are inevitable noise
signals during VIV tests, and there is no generalized processing method for flexible-pipe VIV data.
Therefore, the experimental strain data also contain a noise signal, but they are still the most reliable
benchmark data at present.

{\blueblack 
\cref{uniformres040} illustrates that the response and strain results of the CFD numerical simulation closely align with the experimental results in terms of the symmetric initial IL displacement. Both the VIVANA model and CFD results predict a second-order response with comparable trends in the IL direction, whereas the experimental results show a first-order dominant response. In the CF direction, all the results indicate a first-order dominant response; however, the CFD model underestimates both the VIV response and the corresponding strain. However, it should be noted that this flow case is a type of low-flow-velocity case for the relatively stiff pipe model, which results in first-order dominance retained in the experimental IL direction, as VIV has just begun to occur at this point in the experiment.

Additionally, the underprediction in the IL direction may also be attributed to the limitations of the RANS turbulence model. At low Reynolds numbers, RANS-based simulations may fail to fully resolve unsteady vortex structures that are critical for accurate force estimation, especially in early stage VIV with low flow velocity. Additional comparative studies will be conducted in future work.
}

\begin{figure}[htb!]
    \centering
    \includegraphics[width=.8\textwidth]{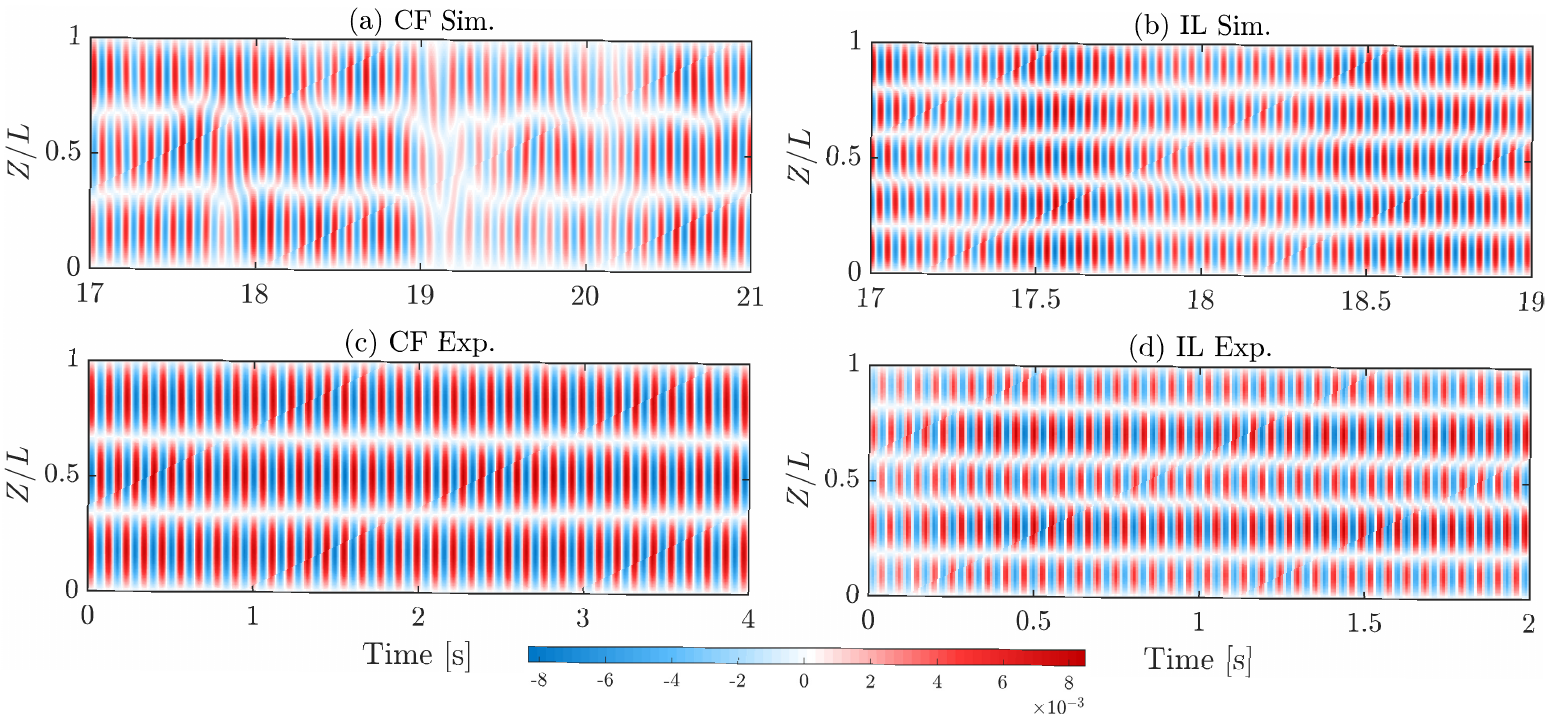}
    \caption{Spatial and temporal distributions of the VIV displacement along the pipe for uniform flow case at $U=\SI{1.60}{m/s}$: (a) and (b) numerical nondimensional displacement (A/D) in the CF and IL directions, respectively; (c) and (d) experimental nondimensional displacement in the CF and IL directions, respectively. Blue to red: negative amplitude to positive amplitude.}
  \label{uniform160travel}
\end{figure}

\cref{uniform040timefre} represents the response frequency at location $Z = 0.2 L$ in the uniform flow $U=\SI{0.40}{m/s}$ case. There are two parts in each subfigure: the time-varying response frequency generated by the wavelet transformation on the left and the general amplitude-frequency spectrum on the right. The continuous wavelet transform equation is
\begin{equation}
\mathcal{WT}_{f}(a, \tau)=\left\langle \xi(t), \psi_{a, \tau}(t)\right\rangle=a^{-1 / 2} \int_{-\infty}^{+\infty} \xi(t) \psi^{*}\left(\frac{t-\tau}{a}\right) d t,
\end{equation}
where $\mathcal{WT}_{f}$ is the wavelet transformation coefficient of the time domain signal $\xi(t)$, which represents the variation in frequency at that time scale. Parameter $a$ is the scale factor, $\tau$ is the shift factor, $\psi(t)$ is the mother wavelet, and the Morlet wavelet is chosen as the mother wavelet. The general amplitude-frequency spectrum is defined as:
\begin{equation}
	\mathcal{F}(f)=\sum_{i=1}^{n} \hat{F}_{i}(f),
\end{equation}
where $\hat{F}_{i}(f)$ is the power spectral density (PSD) at the $i$th data point, and $\mathcal{F}(f)$ is the general amplitude-frequency spectrum obtained by summing the amplitude at the same frequency component for all the data points. This figure shows that the VIV result under these flow conditions exhibits stable first and second order mode responses in the CF and IL directions, respectively, and the numerical simulation in the CF direction presents negligible intermittent responses. The general frequency spectrum clearly has one single dominant peak. The colorbar is omitted in the following text.

\begin{figure}[ht!]
    \centering
    \includegraphics[width=.85\textwidth]{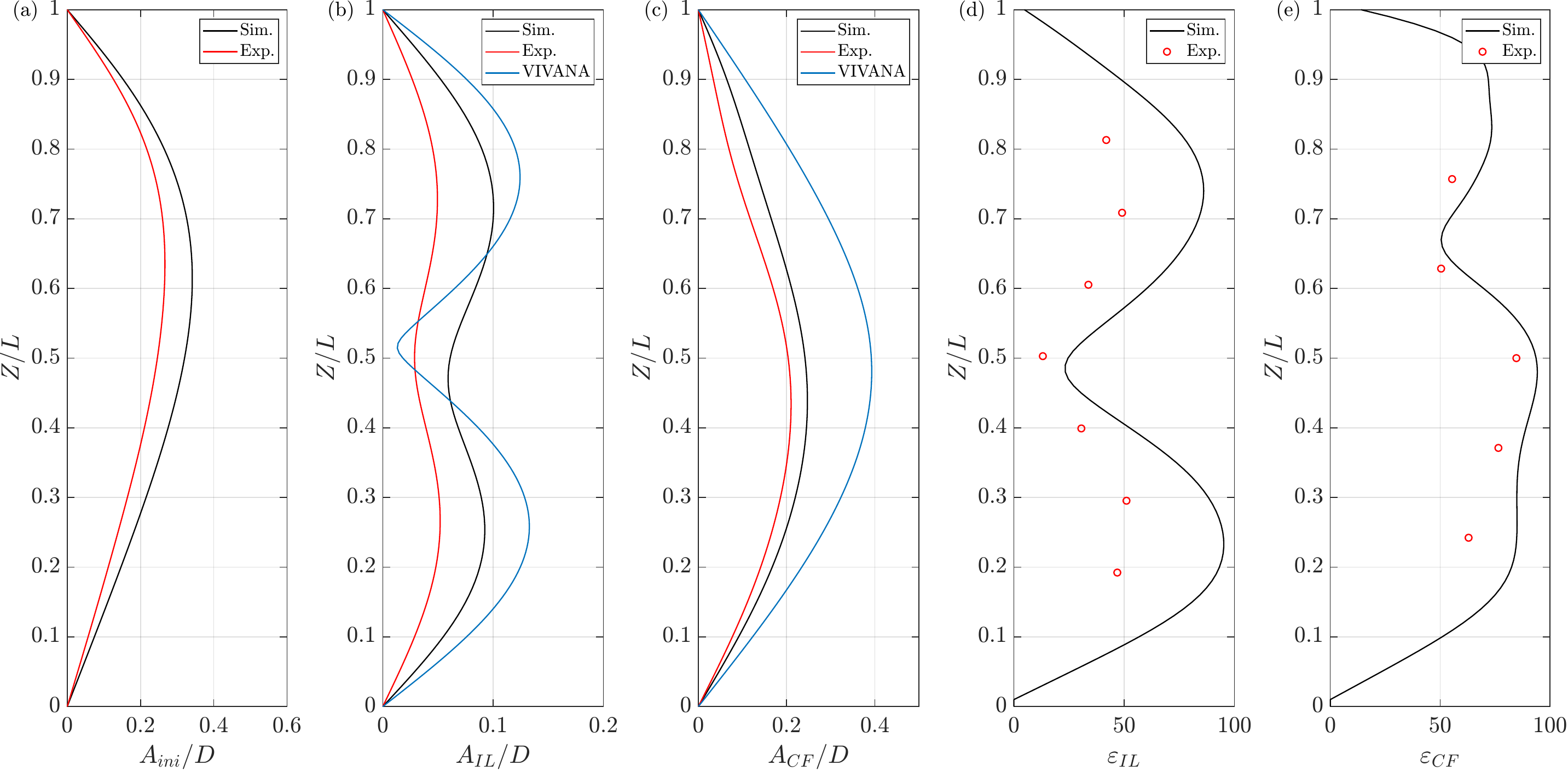}
    \caption{Spanwise distribution of VIV response of linearly sheared flow case at $U_{max}=\SI{0.60}{m/s}$: (a) initial displacement in the IL direction; (b) VIV RMS displacement in the IL direction; (c) VIV RMS displacement in the CF direction; (d) VIV RMS strain in the IL direction; (e) VIV RMS strain in the CF direction. Black line: simulation result; red line and circular symbols: experimental result; blue line: VIVANA result.}
  \label{sf060res}
\end{figure}

\cref{uniformres160} represents the response of $U=\SI{1.60}{m/s}$ case, where the initial IL displacement maintains a symmetric distribution with maximum value at $4D$. The experimental and numerical results are in good agreement, with the IL VIV exhibiting a fifth order response and a third order response in the CF direction. The CFD results in the IL direction align more closely with the experimental data than the VIVANA predictions. Additionally, the strain results from the CFD simulation follow the same trend as the experimental data, demonstrating close agreement.

\begin{figure}[ht!]
    \centering
    \includegraphics[width=.8\textwidth]{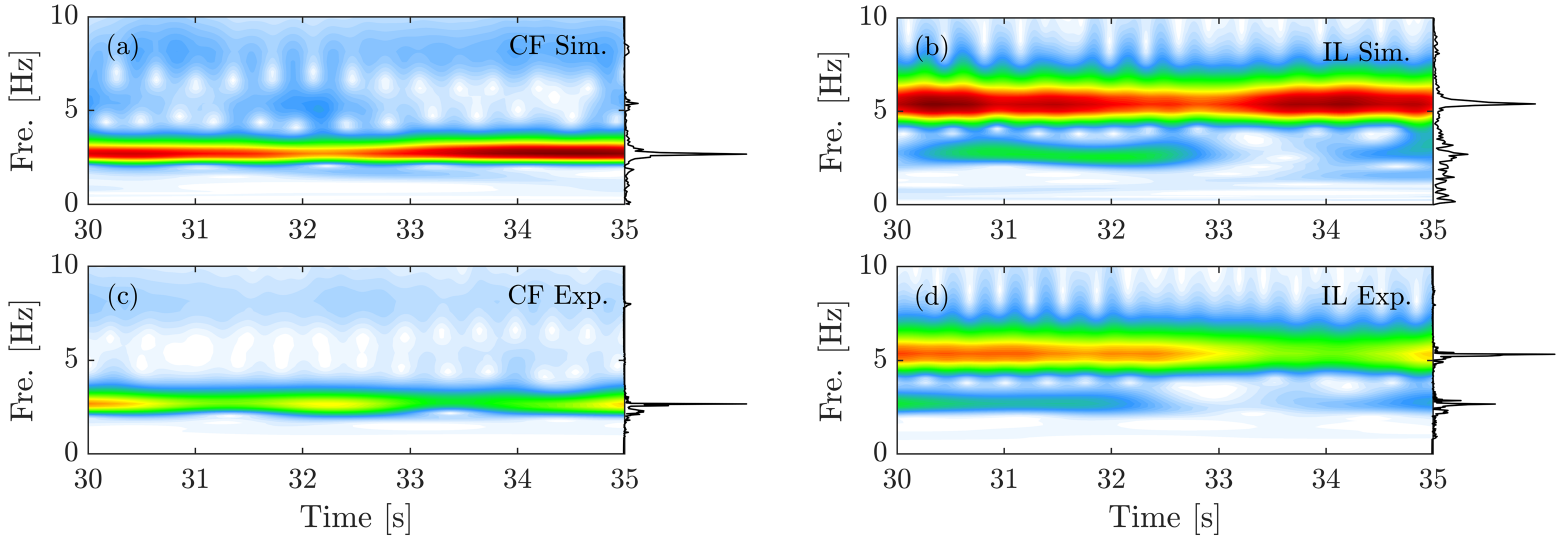}
    \caption{Response frequencies at $Z = 0.2L$ of linearly sheared flow case at $U_{max}=\SI{0.60}{m/s}$ case: (a) time-varying response frequency by wavelet analysis of the numerical result in the CF direction; (b) time-varying response frequency of the numerical result in the IL direction; (c) time-varying response frequency of the experimental result in the CF direction; (d) time-varying response frequency of the experimental result in the IL direction. The black line on the right represents the general frequency spectrum. White to red: zero to maximum value.}
  \label{sf060timefre}
\end{figure}

\cref{uniform160travel} represents the VIV traveling wave phenomenon, and the antinode position variation indicates whether the corresponding VIV response exhibits standing wave or traveling wave characteristics. The results show the vibration response is dominated by the standing wave phenomenon despite a fifth order dominant response in the IL direction. The colorbar is also omitted in the following text.

{\blueblack 
The CFD simulation successfully reproduces the VIV response of a flexible pipe under uniform flow conditions. For low flow velocity ($U = \SI{0.40}{m/s}$) the IL response is first-order dominant in the experiment but second-order in simulation. For the higher velocity case $U = \SI{1.60}{m/s}$, both simulation and experiment show higher mode responses and good agreement in displacement and strain. Traveling wave phenomena are minimal, and the responses are mostly standing wave dominated.}

\begin{figure}[ht!]
    \centering
    \includegraphics[width=.8\textwidth]{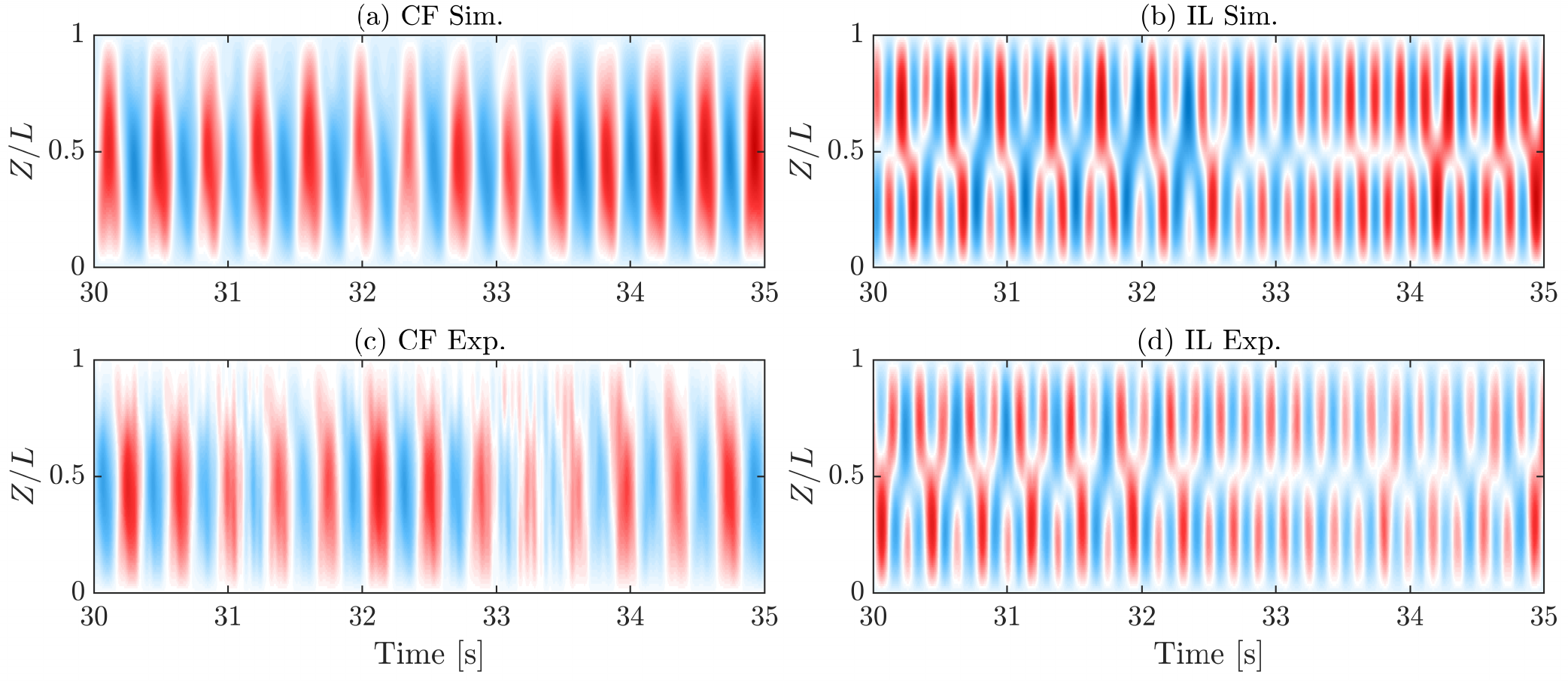}
    \caption{Spatial and temporal distributions of the VIV displacement along the pipe for the linearly sheared flow case at $U_{max}=\SI{0.60}{m/s}$: (a) and (b) numerical nondimensional displacement in the CF and IL directions, respectively; (c) and (d) experimental nondimensional displacement in the CF and IL directions, respectively. Blue to red: negative amplitude to positive amplitude.}
  \label{sf060travel}
\end{figure}

\subsection{Linearly sheared flow cases}
For linearly sheared flow cases, we will introduce two velocity cases as $U_{max} = \SI{0.60}{m/s}$ and $\SI{1.00}{m/s}$. \cref{sf060res} represents the VIV response in the linearly sheared flow case at $U_{max} = \SI{0.60}{m/s}$, and the initial IL displacement exhibits an asymmetric distribution. Due to the sheared flow velocity profile, regions with higher flow velocities have larger initial displacements. The IL and CF VIV responses are dominated by the second order and first order, respectively. The numerical simulation results align more closely with the experimental data than those from VIVANA. 

\begin{figure}[ht!]
    \centering
    \includegraphics[width=.85\textwidth]{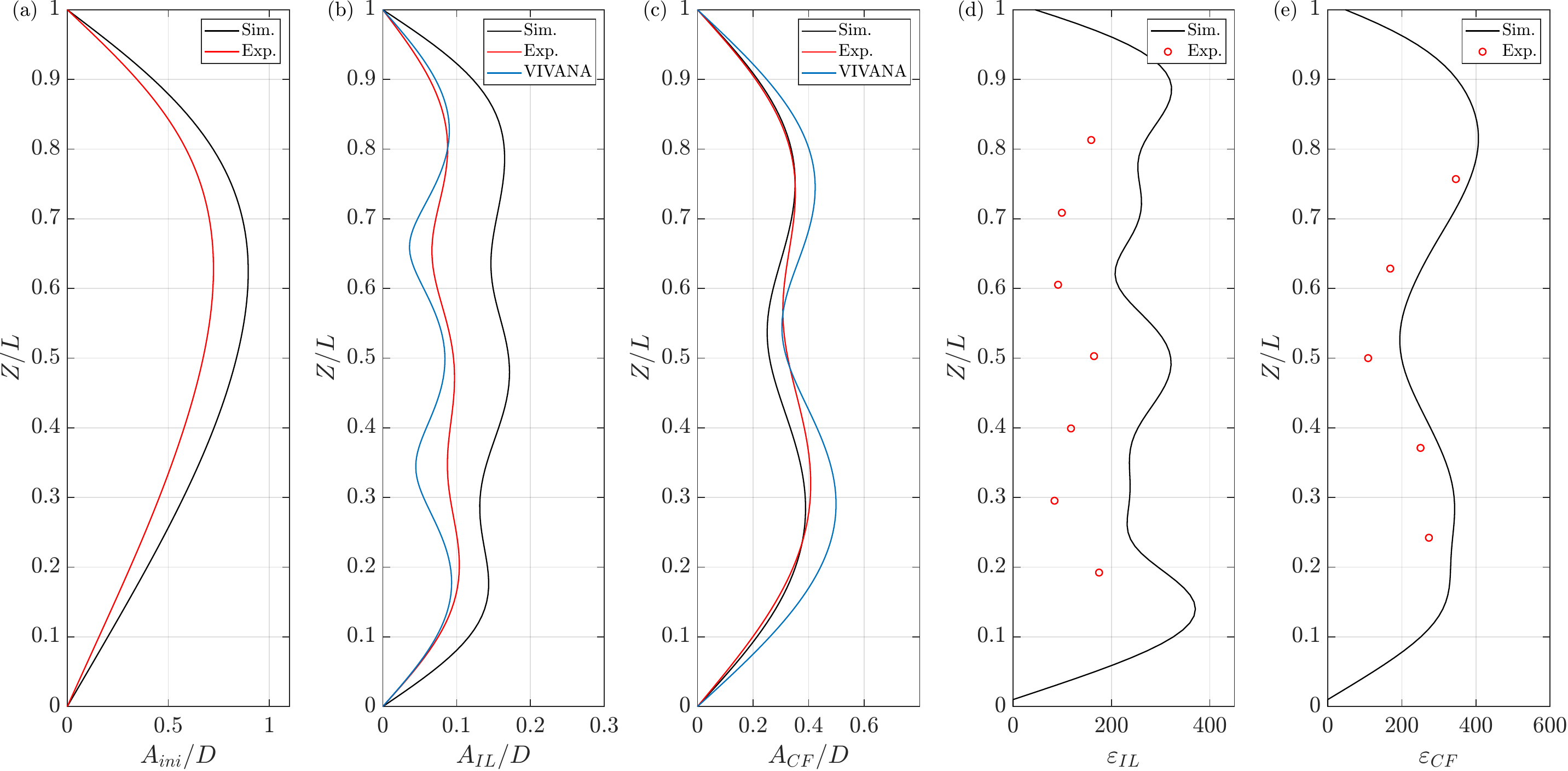}
    \caption{Spanwise distribution of the VIV response of linearly sheared flow case at $U_{max}=\SI{1.00}{m/s}$: (a) initial displacement in the IL direction; (b) VIV RMS displacement in the IL direction; (c) VIV RMS displacement in the CF direction; (d) VIV RMS strain in the IL direction; (e) VIV RMS strain in the CF direction. Black line: simulation result; red line and circular symbol: experimental result; blue line: VIVANA result.}
  \label{sf100res}
\end{figure}

Unlike the uniform flow experiment, which includes 25 strain measurement points in the CF direction and 19 in the IL direction, the linearly sheared flow experiment involves 7 CF strain measurement points and 5 in the IL direction. Owing to the fragility of FBG sensors, in this VIV experiment, the overall displacement is reconstructed via a reduced number of measurement points. In practice, the requirements for experimental modal analysis can be met as long as the number of measurement points exceeds the highest dominant vibration mode \citep{lie2006modal}. The numerical strain results match well with the experimental results.

\begin{figure}[htb!]
    \centering
    \includegraphics[width=.8\textwidth]{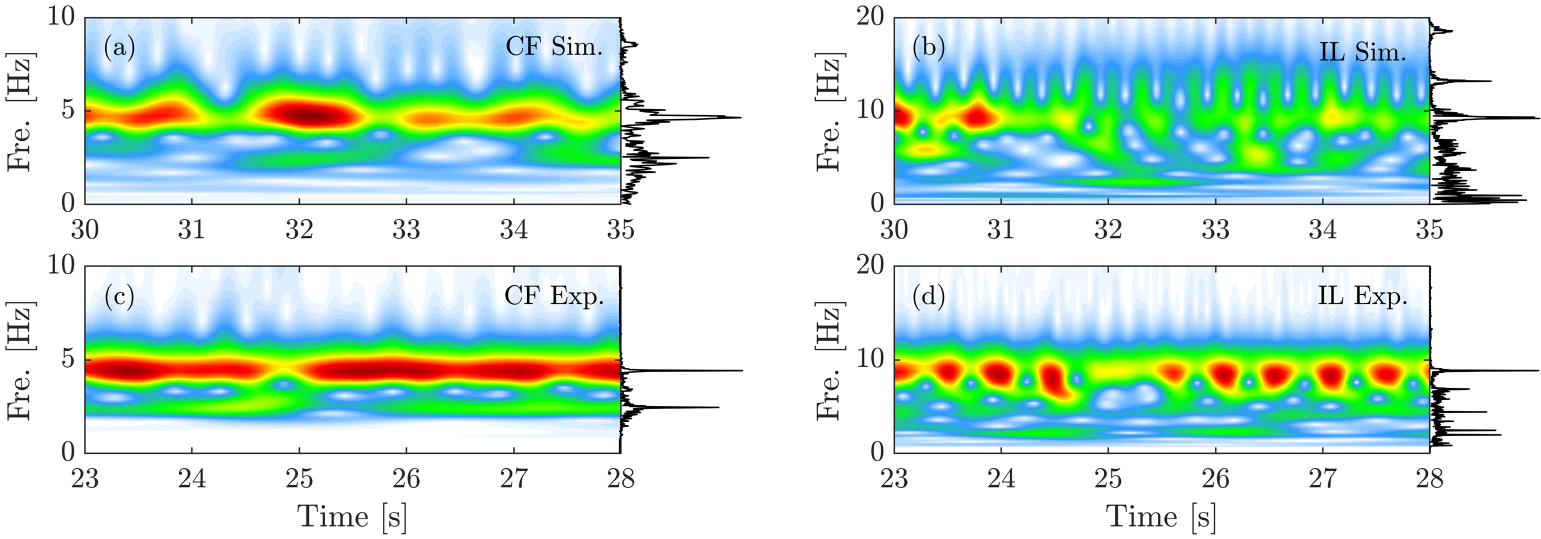}
    \caption{Response frequencies at $Z = 0.2L$ in the linearly sheared flow case at $U_{max}=\SI{1.00}{m/s}$ case: (a) time-varying response frequency via wavelet analysis of the numerical result in the CF direction; (b) time-varying response frequency of the numerical result in the IL direction; (c) time-varying response frequency of the experimental result in the CF direction; (d) time-varying response frequency of the experimental result in the IL direction. Black line on the right represents the general frequency spectrum. White to red: zero to maximum value.}
  \label{sf100timefre}
\end{figure}

Unlike the VIVANA prediction results, the IL and CF responses contain multiple first-order components. As \cref{sf060timefre} shows, the VIV experiments generally exhibit a time-sharing phenomenon, where the dominant frequency in the wavelet result remains singular over time \citep{swithenbank2012occurrence}. This is one basic assumption in current engineering prediction models such as VIVANA. However, experimental data often contain additional harmonic components that do not dominate the vibration but result in differences in the prediction results, as shown in subfigures~(b) and (d).

\begin{figure}[htb!]
    \centering
    \includegraphics[width=.8\textwidth]{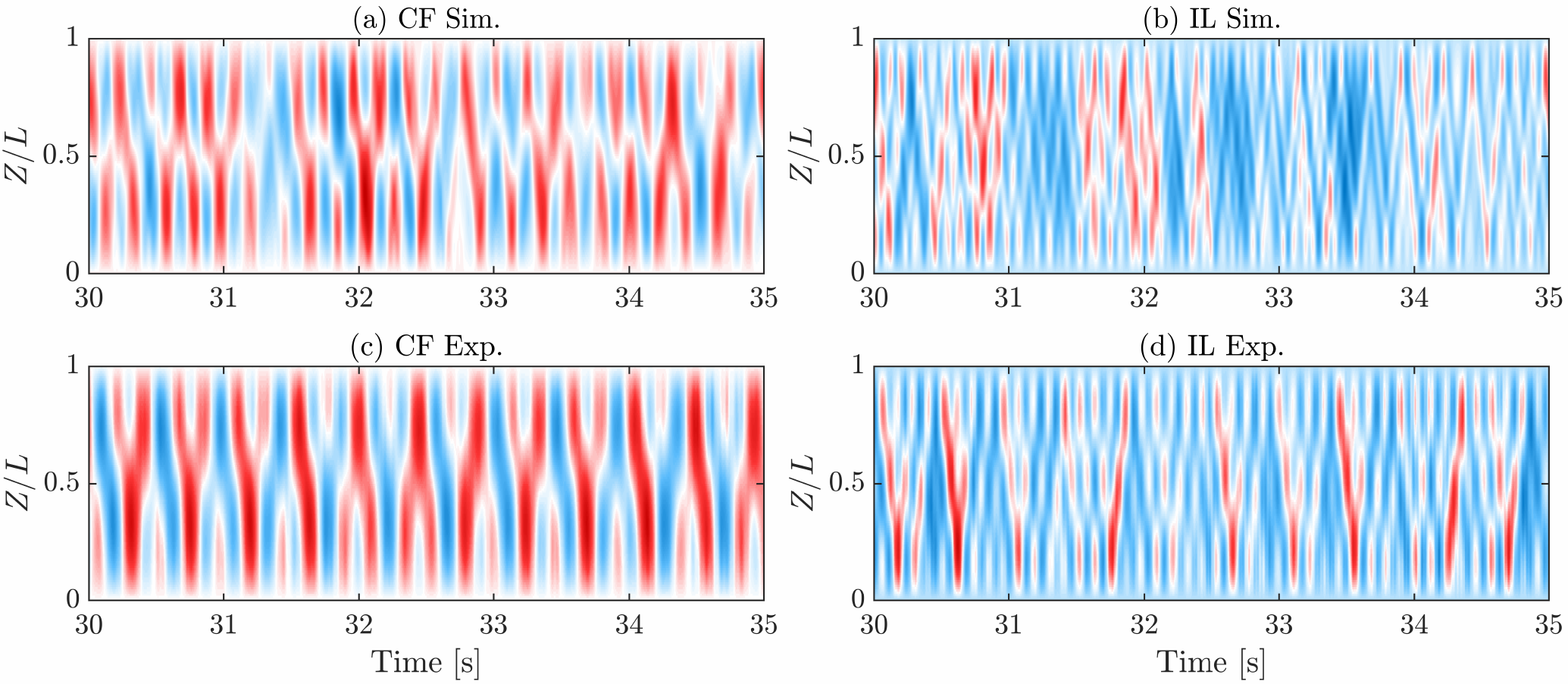}
    \caption{Spatial and temporal distributions of the VIV displacement along the pipe for the linearly sheared flow case at $U_{max}=\SI{1.00}{m/s}$: (a) and (b) numerical nondimensional displacement (A/D) in the CF and IL directions, respectively; (c) and (d) experimental nondimensional displacement in the CF and IL directions, respectively. Blue to red: negative amplitude to positive amplitude.}
  \label{sf100travel}
\end{figure}

\cref{sf100res} represents the response result of linearly sheared flow case $U_{max} = \SI{1.00}{m/s}$. Similar to the previous case, the initial IL displacement exhibits an asymmetric distribution with the third order dominant response in the IL direction and second order dominant response in the CF direction. Meanwhile, in this flow case, both the VIVANA predictions and the numerical simulations in this study show good agreement with the experimental results. Moreover, the strain results from the numerical simulation closely match the experimental data in both trend and magnitude. The numerical simulation results are higher than the experimental results.

\cref{sf100res} represents the response result in the linearly sheared flow case $U_{max} = \SI{1.00}{m/s}$. Similar to the previous case, in this case, the initial IL displacement has an asymmetric distribution, with a third-order dominant response in the IL direction and a second-order dominant response in the CF direction. Moreover, in this flow case, both the VIVANA predictions and the numerical simulations show good agreement with the experimental results. Moreover, the strain results from the numerical simulation closely match the experimental data in both trend and magnitude. The numerical simulation results are better than the experimental results.

{\blueblack
The VIV response under linearly sheared flow shows increased complexity compared to uniform flow. Asymmetry in initial IL displacement is observed due to the non-uniform velocity profile. Numerical simulations better match the experimental data than VIVANA in few cases. Traveling wave features begin to emerge, and time–frequency analysis reveals time-sharing behavior. }

\subsection{Bidirectionally sheared flow cases}
In this flow case, we introduce three flow velocity cases: $U_{max} = \SI{0.48}{m/s}$,  $\SI{0.77}{m/s}$ and $\SI{0.99}{m/s}$.

The two flow conditions discussed above have been widely studied in academia and engineering fields, with established engineering prediction frameworks available, reinforcing the credibility of the numerical simulation results presented in this study. However, bidirectionally sheared flow was first experimentally investigated in a laboratory in 2022, revealing several phenomena that deviate from the prevailing understanding of vortex-induced vibration. Given the large experimental scale and the impracticality of directly observing the flow field, the numerical simulation scheme developed in this study could serve as a tool for future fluid mechanism investigations. Nevertheless, owing to length constraints, this paper focuses primarily on a comparative study of the VIV response under the discussed flow conditions.
{\blueblack
\begin{figure}[htb!]
    \centering
    \includegraphics[width=.85\textwidth]{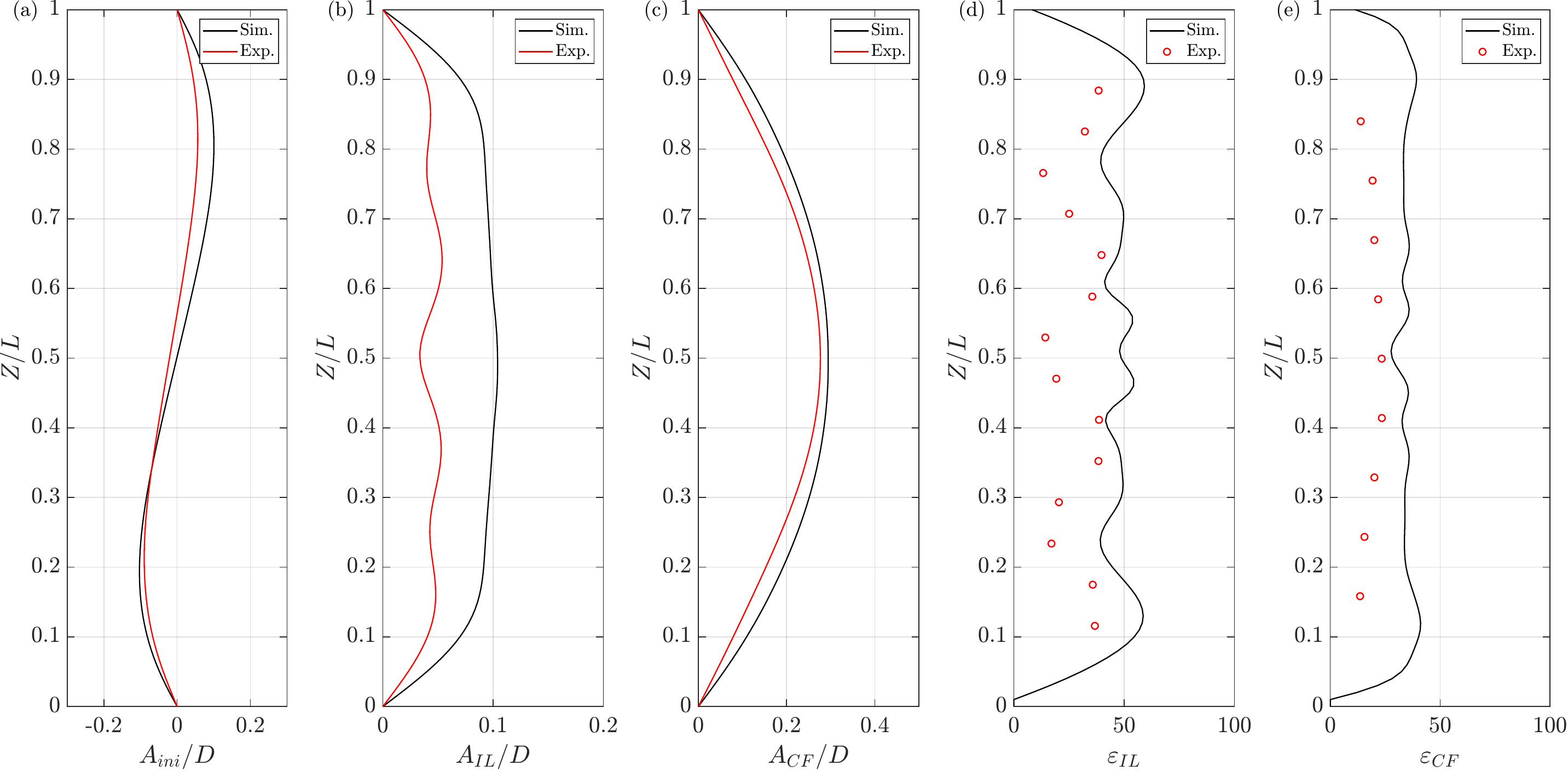}
    \caption{Spanwise distribution of VIV response in the bidirectionally sheared flow case at $U_{max}=\SI{0.48}{m/s}$: (a) initial displacement in the IL direction; (b) VIV RMS displacement in the IL direction; (c) VIV RMS displacement in the CF direction; (d) VIV RMS strain in the IL direction; (e) VIV RMS strain in the CF direction. Black line: simulation result; red line and circular symbol: experimental result.}
  \label{bi048res}
\end{figure}

\cref{bi048res} represents the VIV response in the bidirectionally sheared flow case at $U_{max} = \SI{0.48}{m/s}$. There is currently no existing engineering VIV prediction framework for this flow field VIV now. The initial IL displacement displays an antisymmetric distribution with a second-order modal shape. We need to analyze the IL and CF direction responses by combining the results of time-frequency analyses as \cref{bi048timefre} shown.

\begin{figure}[htb!]
    \centering
    \includegraphics[width=.8\textwidth]{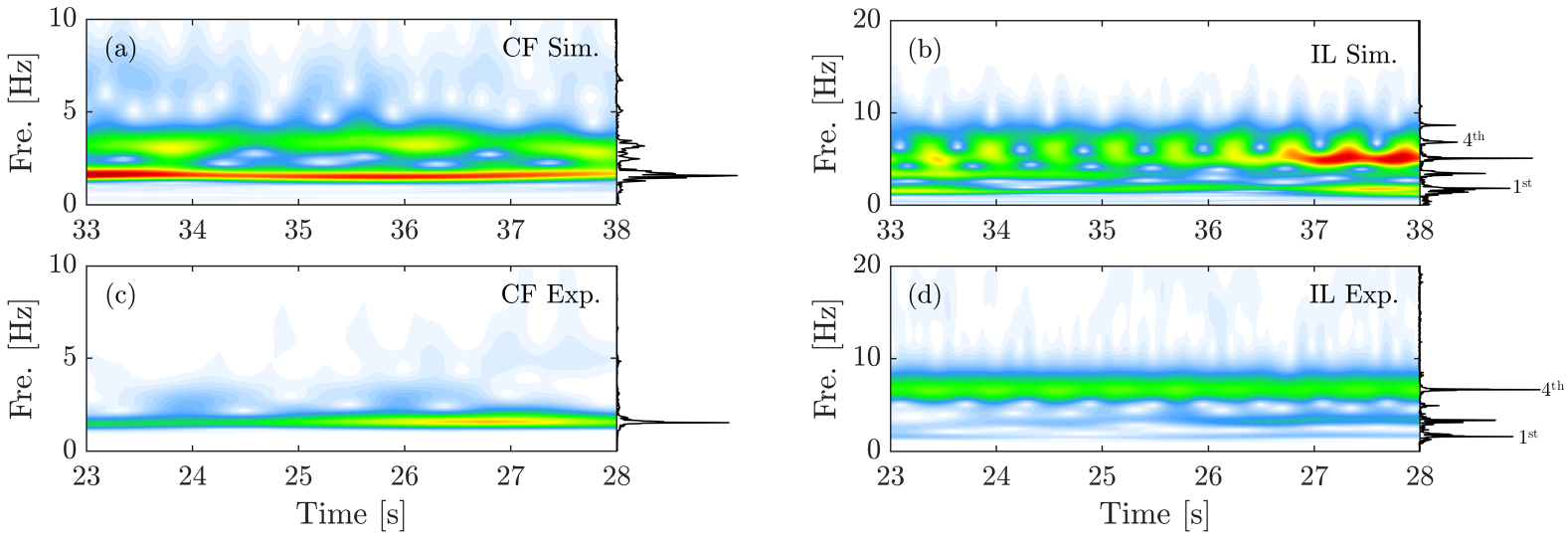}
    \caption{Response frequencies at $Z = 0.2L$ of bidirectionally sheared flow case at $U_{max}=\SI{0.48}{m/s}$: (a) time-varying response frequency obtained via wavelet analysis of the numerical result in the CF direction; (b) time-varying response frequency of the numerical result in the IL direction; (c) time-varying response frequency of the experimental result in the CF direction; (d) time-varying response frequency of the experimental result in the IL direction. Black line on the right represents the general frequency spectrum. White to red: zero to maximum value.}
  \label{bi048timefre}
\end{figure}
 }
Notably, this flow velocity is relatively low, with a first order dominant CF response at nearly $0.3D$, but a fairly complex response is observed in the IL direction. In the general frequency spectrum, a first to fifth order response can be observed, whereas a fourth order response is shown in the experimental result and a third-order response is present in the simulation result.

\begin{figure}[htb!]
    \centering
    \includegraphics[width=.8\textwidth]{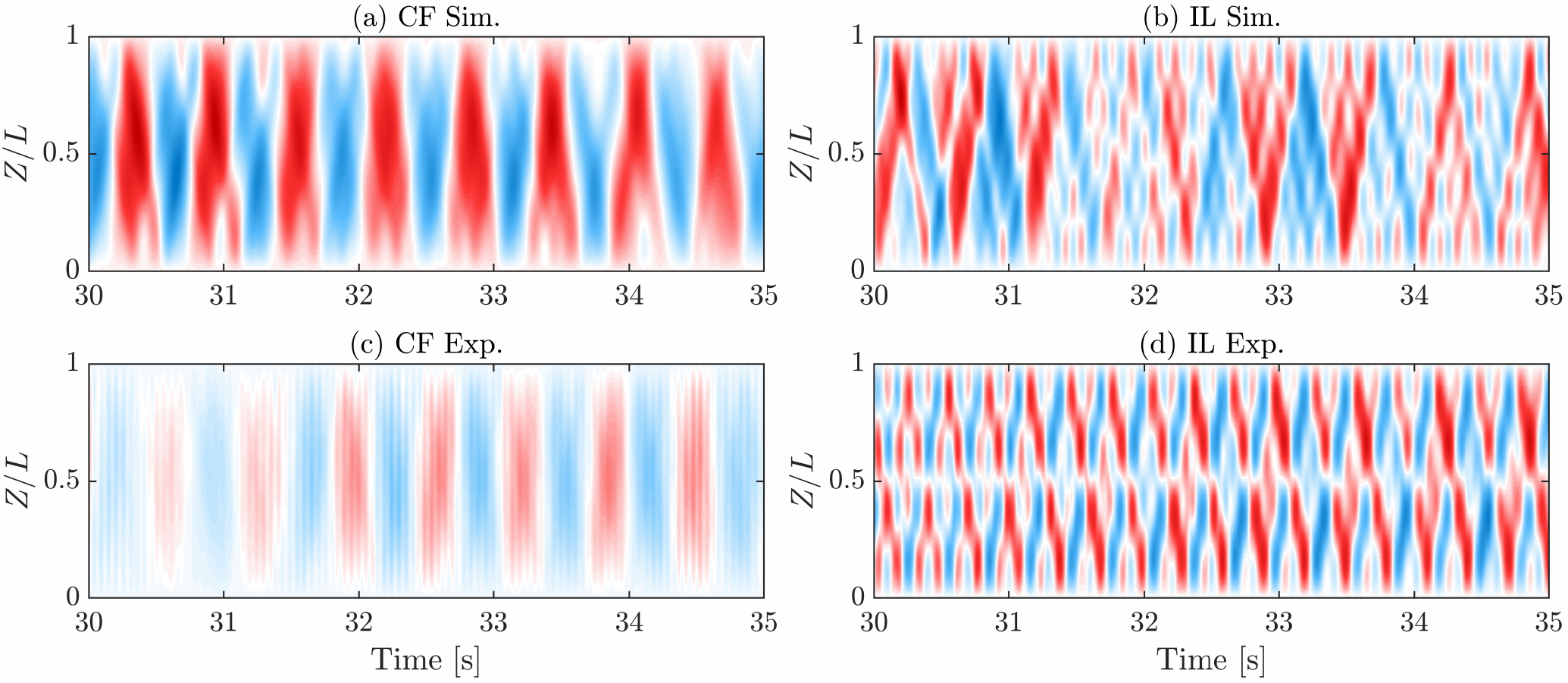}
    \caption{Spatial and temporal distributions of the VIV displacement along the pipe of bidirectionally sheared flow case at $U_{max}=\SI{0.48}{m/s}$: (a) and (b) numerical nondimensional displacement (A/D) in the CF and IL directions, respectively; (c) and (d) experimental nondimensional displacement in the CF and IL directions, respectively. Blue to red: negative amplitude to positive amplitude.}
  \label{bi048travel}
\end{figure}

The common feature is that under such a low flow velocity condition, there is a distinct multifrequency response, but the dominant mode is different. Notably, in the numerical simulation method used in this study, constant tension is assumed during VIV. However, in the actual experiment, the tensioners at the end of the system apply variable tension through springs. Structural nonlinearity was not incorporated in the present numerical simulation. In our previous VIV engineering prediction studies, where variable tension was considered, we found that the experimental riser, which is composed of multiple layers, exhibits a variable axial stiffness under vortex-induced vibration, which does not always align with the material test value. We are currently developing a more accurate model to account for the effects of variable tension in future studies. Moreover, the traveling wave response also has a significant effect on this multi-frequency response. We confirm that the multifrequency IL phenomenon under these flow velocity conditions is real rather than an experimental error.

\begin{figure}[htb!]
    \centering
    \includegraphics[width=.85\textwidth]{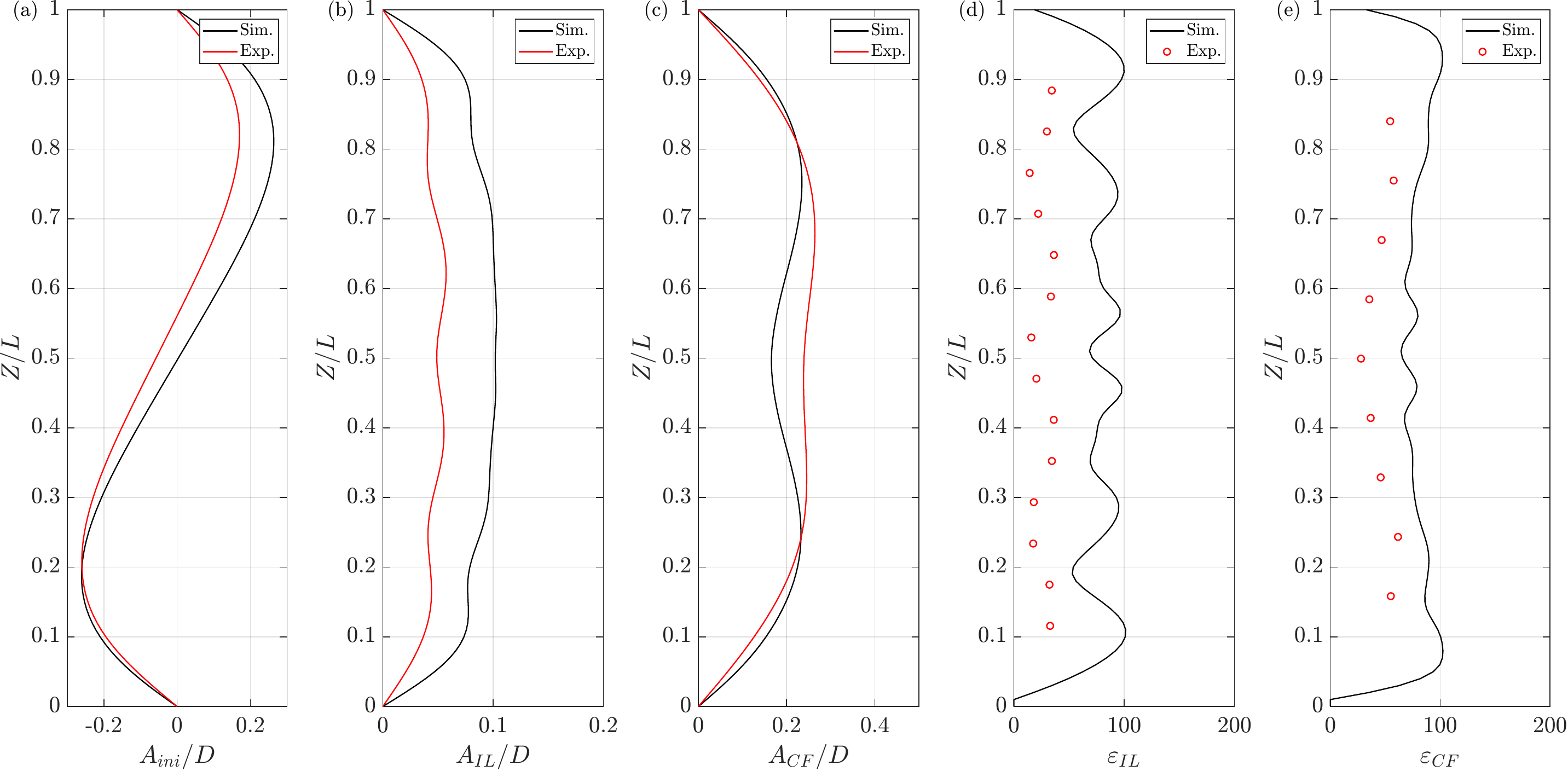}
    \caption{Spanwise distribution of VIV response of bidirectionally sheared flow case at $U_{max}=\SI{0.77}{m/s}$: (a) initial displacement in the IL direction; (b) VIV RMS displacement in the IL direction; (c) VIV RMS displacement in the CF direction; (d) VIV RMS strain in the IL direction; (e) VIV RMS strain in the CF direction. Black line: simulation result; red line and circular symbols: experimental result.}
  \label{bi077res}
\end{figure}

\cref{bi048travel} represents the spatial and temporal distributions of the VIV response. There exists an obvious traveling phenomenon in the IL direction with high order response, whereas the CF response is first order dominated.
 
\begin{figure}[htb!]
    \centering
    \includegraphics[width=.8\textwidth]{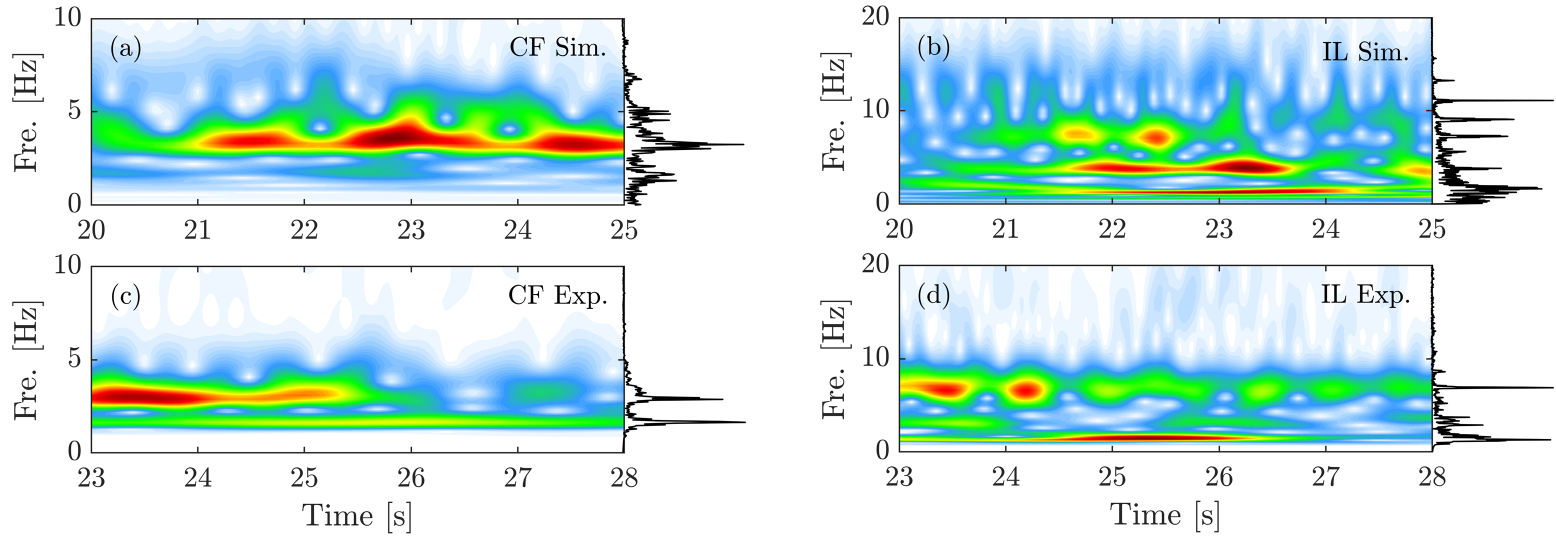}
    \caption{Response frequencies at $Z = 0.2L$ of bidirectionally sheared flow case at $U_{max}=\SI{0.77}{m/s}$: (a) time-varying response frequency by wavelet analysis of numerical result in the CF direction; (b) time-varying response frequency of numerical result in the IL direction; (c) time-varying response frequency of experimental result in the CF direction; (d) time-varying response frequency of experimental result in the IL direction. Black line on the right represents the general frequency spectrum. White to red: zero to maximum value.}
  \label{bi077timefre}
\end{figure}

\cref{bi077res} and \cref{bi077timefre} represent the VIV response of the bidirectionally sheared case of $U_{max} = \SI{0.77}{m/s}$. In this flow case, the VIV response in the CF direction represents multi-frequency of the first order and the second order, the numerical result and experimental result are close. A complex multi-frequency response is also observed in the IL direction. Compared with the experimental results, the numerical simulation results in this study consistently yield higher strain
values.

\begin{figure}[htb!]
    \centering
    \includegraphics[width=.8\textwidth]{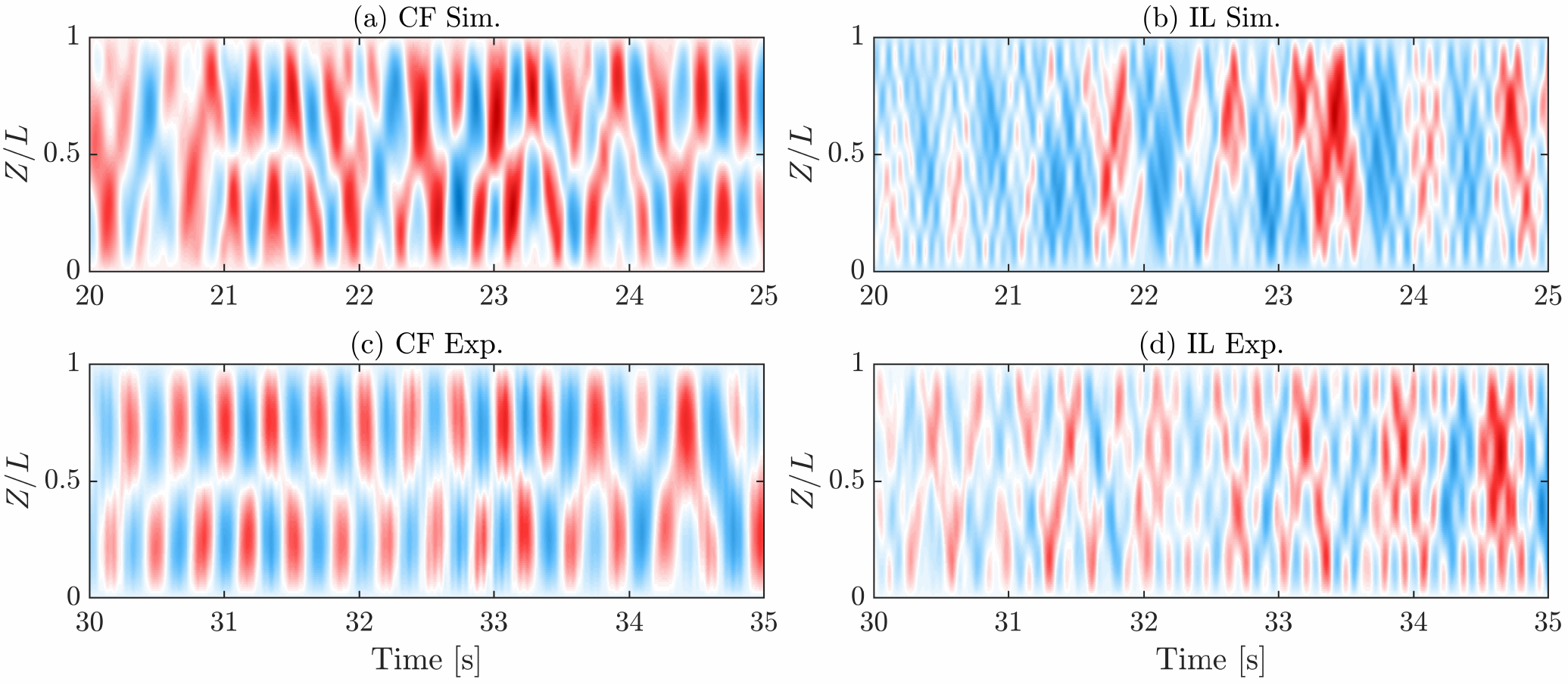}
    \caption{Spatial and temporal distributions of the VIV displacement along the pipe of bidirectionally sheared flow case at $U_{max}=\SI{0.77}{m/s}$: (a) and (b) numerical nondimensional displacement (A/D) in the CF and IL directions, respectively; (c) and (d) experimental nondimensional displacement in the CF and IL directions, respectively. Blue to red: negative amplitude to positive amplitude.}
  \label{bi077travel}
\end{figure}

\begin{figure}[htb!]
    \centering
    \includegraphics[width=.85\textwidth]{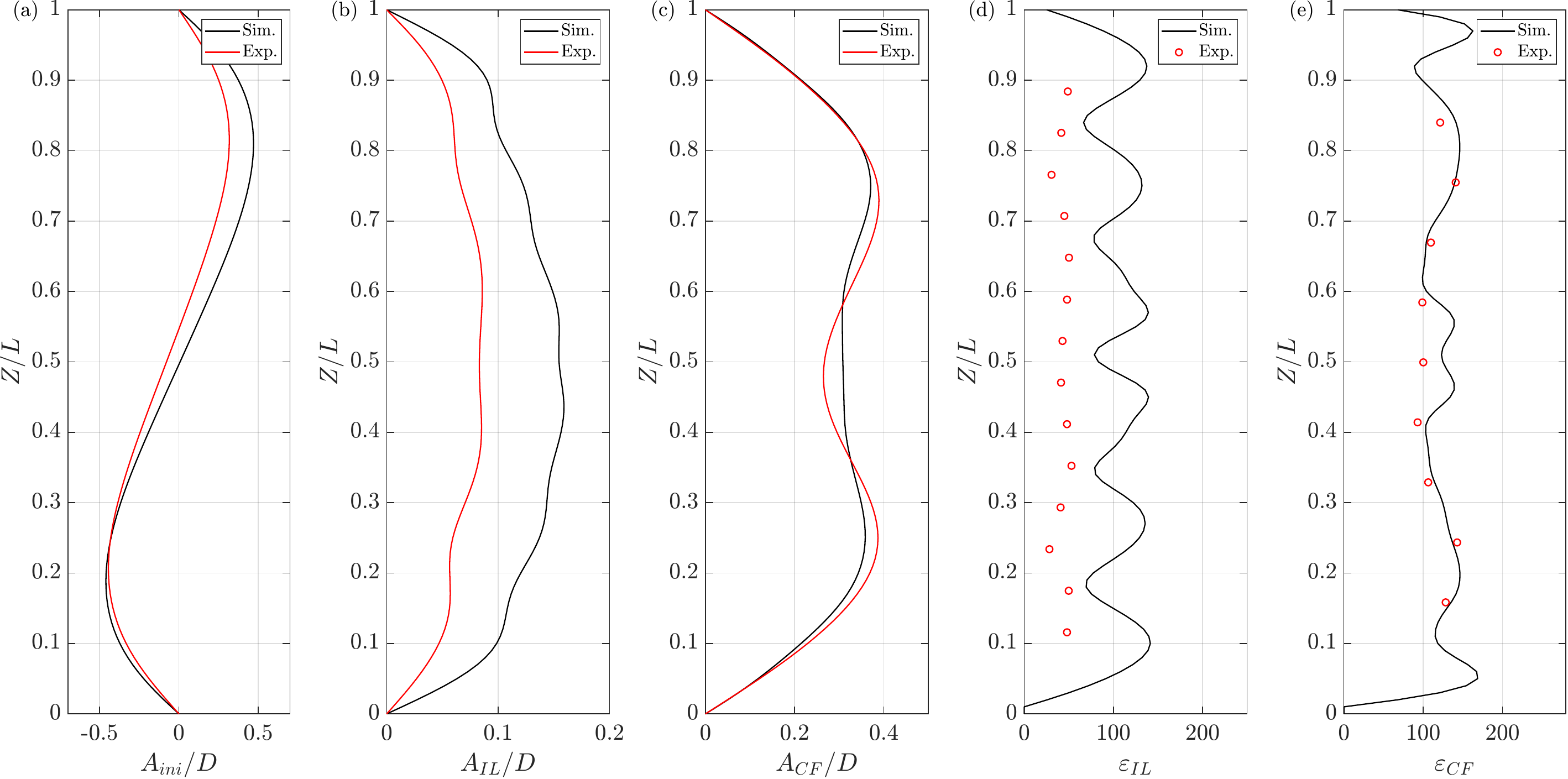}
    \caption{Spanwise distribution of VIV response of bidirectionally sheared flow case at $U_{max}=\SI{0.99}{m/s}$: (a) initial displacement in the IL direction; (b) VIV RMS displacement in the IL direction; (c) VIV RMS displacement in the CF direction; (d) VIV RMS strain in the IL direction; (e) VIV RMS strain in the CF direction. Black line: simulation result; red line and circular symbols: experimental result.}
  \label{bi099res}
\end{figure}

\begin{figure}[htb!]
    \centering
    \includegraphics[width=.8\textwidth]{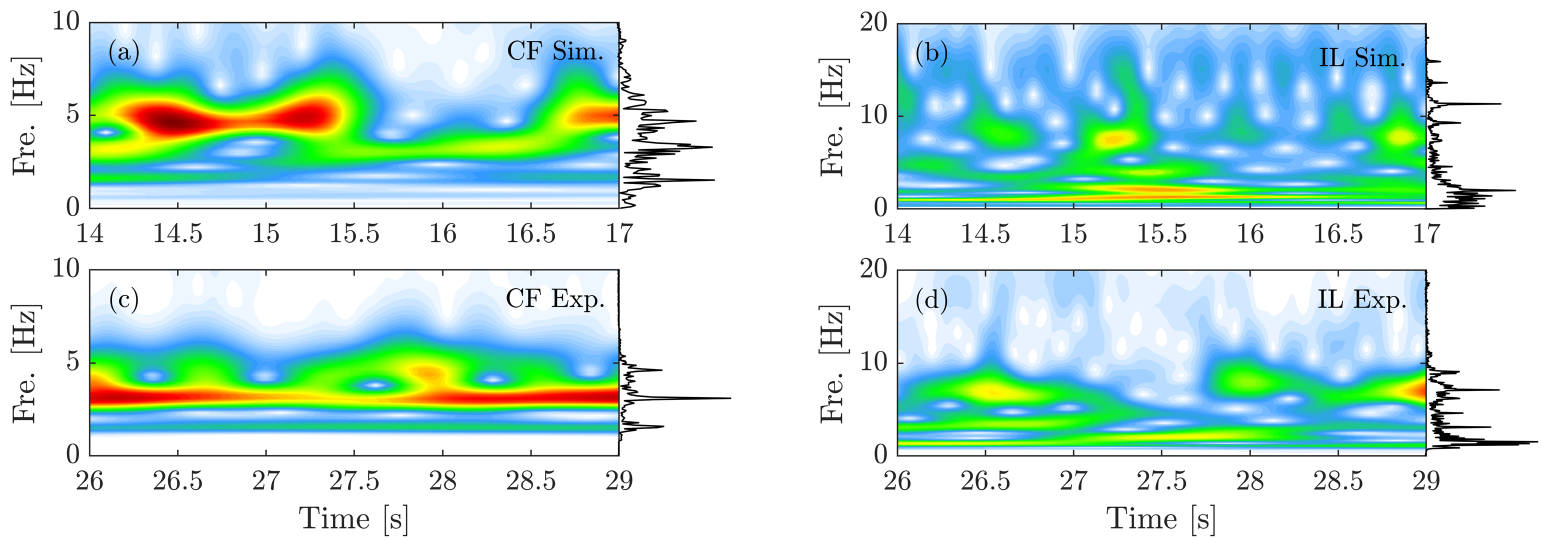}
    \caption{Response frequencies at $Z = 0.2L$ of bidirectionally sheared flow case at $U_{max}=\SI{0.99}{m/s}$: (a) time-varying response frequency by wavelet analysis of the numerical result in the CF direction; (b) time-varying response frequency of the numerical result in the IL direction; (c) time-varying response frequency of the experimental result in the CF direction; (d) time-varying response frequency of the experimental result in the IL direction. Black line on the right represents the general frequency spectrum. White to red: zero to maximum value.}
  \label{bi099timefre}
\end{figure}

\begin{figure}[htb!]
    \centering
    \includegraphics[width=.8\textwidth]{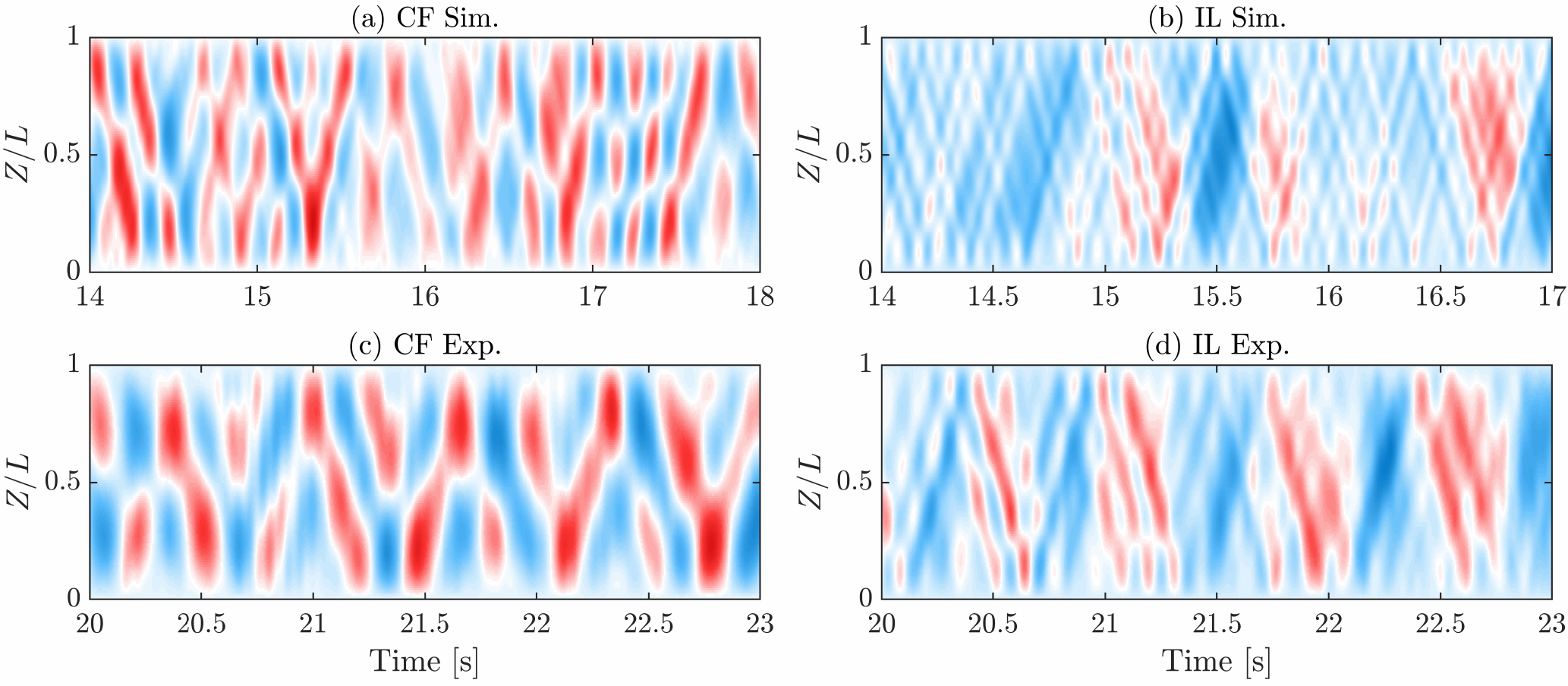}
    \caption{Spatial and temporal distributions of the VIV displacement along the pipe of bidirectionally sheared flow case at $U_{max}=\SI{0.99}{m/s}$: (a) and (b) numerical nondimensional displacement (A/D) in the CF and IL directions, respectively; (c) and (d) experimental nondimensional displacement in the CF and IL directions, respectively. Blue to red: negative amplitude to positive amplitude.}
  \label{bi099travel}
\end{figure}

\cref{bi077travel} represents the obvious traveling phenomenon in both the CF and the IL directions, with a low second-order response in the CF direction, which is a typical phenomenon observed experimentally. \cref{bi099res} to \cref{bi099travel} illustrate the VIV response results in the bidirectionally sheared flow case at $U_{max}=\SI{0.99}{m/s}$. Similarly, an antisymmeric initial IL response exists, with the VIV response in the CF direction displaying second-order dominance and that in the IL direction exhibiting considerable complexity. 

The numerical simulation results in the CF direction are in good agreement with the experimental results in terms of both response and strain. However, the numerical simulation results in the IL direction are higher than the experimental results. The initial IL displacement in the numerical simulation is more symmetric than that in the experimental results. This discrepancy likely arises from the difficulty in maintaining a consistently perfect antisymmetric field around the pipe model during the experiment.

\cref{bi099res} to \cref{bi099travel} represent the VIV response in the bidirectionally sheared flow case at $U_{max}=\SI{0.99}{m/s}$. The initial IL displacement maintains an antisymmetric distribution, exhibiting a clear multifrequency response in the IL direction. Moreover, a second-order response is primarily dominant in the CF direction, with third-order contributions. In the bidirectionally sheared flow simulation, the numerical results in the IL direction consistently exceed the experimental values, whereas the results in the CF direction closely align with the experimental data. The strain results follow a similar pattern. During the experiment, the IL VIV response always contains an obvious first-order response with a high-order component, which is also found in the simulation results. The numerical simulations in this study capture this phenomenon, and more detailed hydrodynamic investigations will be conducted in the future.  A distinct multifrequency response under these flow velocity conditions is also observed in the CF direction, as \cref{bi099timefre} (a) and (c) show, with a distinct traveling wave response in both the CF and the IL directions.  

{\blueblack Under bidirectionally sheared flow, the VIV response becomes significantly more complex. Simulations reveal multi-frequency responses and traveling waves in both IL and CF directions. While experimental and numerical results agree qualitatively, discrepancies in IL amplitude are observed, possibly due to structural nonlinearity and variable tension in the experiment. This marks the first successful numerical capture of these phenomena, also highlighting the need for improved physical modeling of such complex flows as future work.
}

{\blueblack 
\section{Conclusions and future work}
In this study, a validated numerical framework for predicting VIV of a flexible pipe in steady flows is developed. It integrates URANS-based fluid dynamics with a structural finite element model through a weak coupling scheme. The model is systematically validated against three experimental benchmark cases: uniform, linearly sheared, and bidirectionally sheared flows. The simulation captures key features of the VIV response, including displacement, frequency, and strain, with agreement with experimental results. Compared to frequency-domain tools, it shows improved accuracy in reproducing high-order modal responses and traveling wave phenomena, particularly under shear flow conditions. Notably, this study presents the first successful numerical simulation of flexible pipe VIV under bidirectionally sheared flow, capturing antisymmetric initial deformation, multifrequency in-line responses, and strong traveling wave features as observed in experiments.

However, the model has the following limitations with future work requirements: 1) The Euler-Bernoulli beam theory is applied for the structural model, which may not be sufficient for the more nonlinear beam models required for structures such as steel catenary risers. 2) The model does not account for the time-domain variation in tension, which introduces some errors in the simulation of high-order responses. 3) For multipipe systems, it remains unclear whether URANS can accurately capture the wake field and the potential problem with possible reverse energy cascades for 2D LES strips \citep{chen2006physical}. The applicability of the strip method to multipipe systems will be studied further in the future. 4) The current study focuses primarily on structural responses. Further investigations into hydrodynamic aspects—such as wake patterns, VIV force identification, and their relationship to excitation and added mass coefficients, will be conducted to enhance the understanding of VIV mechanism.  }

\section*{Data availability}
The corresponding benchmark experimental data and VIV simulation code can be found in the GitHub repository: \url{https://github.com/xuepengfu/VIVdatashare}. There are also additional case
results in this repository. The program and experimental data will be made publicly available as the research progresses.

\section*{Acknowledgements}
The authors gratefully acknowledge the National Science Fund for Distinguished Young Scholars (Grant No.52425102).

\appendix
\section{Finite element model of a tensioned flexible pipe}\label{appenfem}
The two degrees of freedom beam model (IL and CF directions) are applied in the present study. The element mass matrix is given by:
\begin{equation}
    \boldsymbol{M}^e=\frac{\bar{m} l}{420}\left[\begin{array}{cccc}
156 & 22 l & 54 & -13 l \\
22 l & 4 l^2 & 13 l & -3 l^2 \\
54 & 13 l & 156 & -22 l \\
-13 l & -3 l^2 & -22 l & 4 l^2
\end{array}\right],
\end{equation}
and the stiffness matrix is:
\begin{equation}
\boldsymbol{K}^e=\boldsymbol{K}^E+\boldsymbol{K}^T,
\end{equation}
where
\begin{equation}
 \boldsymbol{K}^E = \frac{E I}{l^3}\left[\begin{array}{cccc}
12 & 6 l & -12 & 6 l \\
6 l & 4 l^2 & -6 l & 2 l^2 \\
-12 & -6 l & 12 & -6 l \\
6 l & 2 l^2 & -6 l & 4 l^2
\end{array}\right] ,
 \boldsymbol{K}^T = \frac{T}{30 l}\left[\begin{array}{cccc}
36 & 3 l & -36 & 3 l \\
3 l & 4 l^2 & -3 l & -l^2 \\
-36 & -3 l & 36 & -3 l \\
3 l & -l^2 & -3 l & 4 l^2
\end{array}\right],
\end{equation}
where Rayleigh damping is used for the damping matrix, expressed as:
\begin{equation}
\begin{aligned}
&\boldsymbol{C}=\alpha\boldsymbol{M}+\beta\boldsymbol{K}, \\
&\alpha=\frac{2(\zeta_{1}\omega_{2}-\zeta_{2}\omega_{1})}{(\omega_{1}+\omega_{2})(\omega_{2}-\omega_{1})}\omega_{1}\omega_{2}, \\
&\beta =\frac{2(\zeta_1\omega_2-\zeta_2\omega_1)}{(\omega_1+\omega_2)(\omega_2-\omega_1)},
\end{aligned}
\end{equation}
where $ \bar{m} $ is the unit mass, $ l $ is the length of the riser, $ EI $ is the bending stiffness, $ T $ is the tension, $ \omega_{1} $ and $ \omega_{2} $ are the first two natural circular frequencies of the pipe, and $ \zeta_{1} $ and $ \zeta_{2} $ are the first two damping ratios.

The system mass and stiffness matrices are obtained by assembling the element mass and stiffness matrices as:
\begin{equation}\boldsymbol{K} = 
\left[
\begin{array}{cccccc}
\boldsymbol{K}^e_{(1.1)} & \cdots & 0 & \cdots & 0 \\
 & \ddots & \vdots & \ddots & \vdots \\
 & & \boldsymbol{K}^e_{(i.2)} + \boldsymbol{K}^e_{((i+1).1)} & \cdots & 0 \\
 &\text{symmetric} &  & \ddots & \vdots \\
 & & & & \boldsymbol{K}^e_{(N.2)}
\end{array}
\right]
\end{equation}
\begin{equation}\boldsymbol{M} = 
\left[
\begin{array}{cccccc}
\boldsymbol{M}^e_{(1.1)} & \cdots & 0 & \cdots & 0 \\
 & \ddots & \vdots & \ddots & \vdots \\
 & & \boldsymbol{M}^e_{(i.2)} + \boldsymbol{M}^e_{((i+1).1)} & \cdots & 0 \\
 &\text{symmetric} &  & \ddots & \vdots \\
 & & & & \boldsymbol{M}^e_{(N.2)}
\end{array}
\right]
\end{equation}

\section{Independence test}
\begin{figure}[htb!]
    \centering
    \includegraphics[width=.5\textwidth]{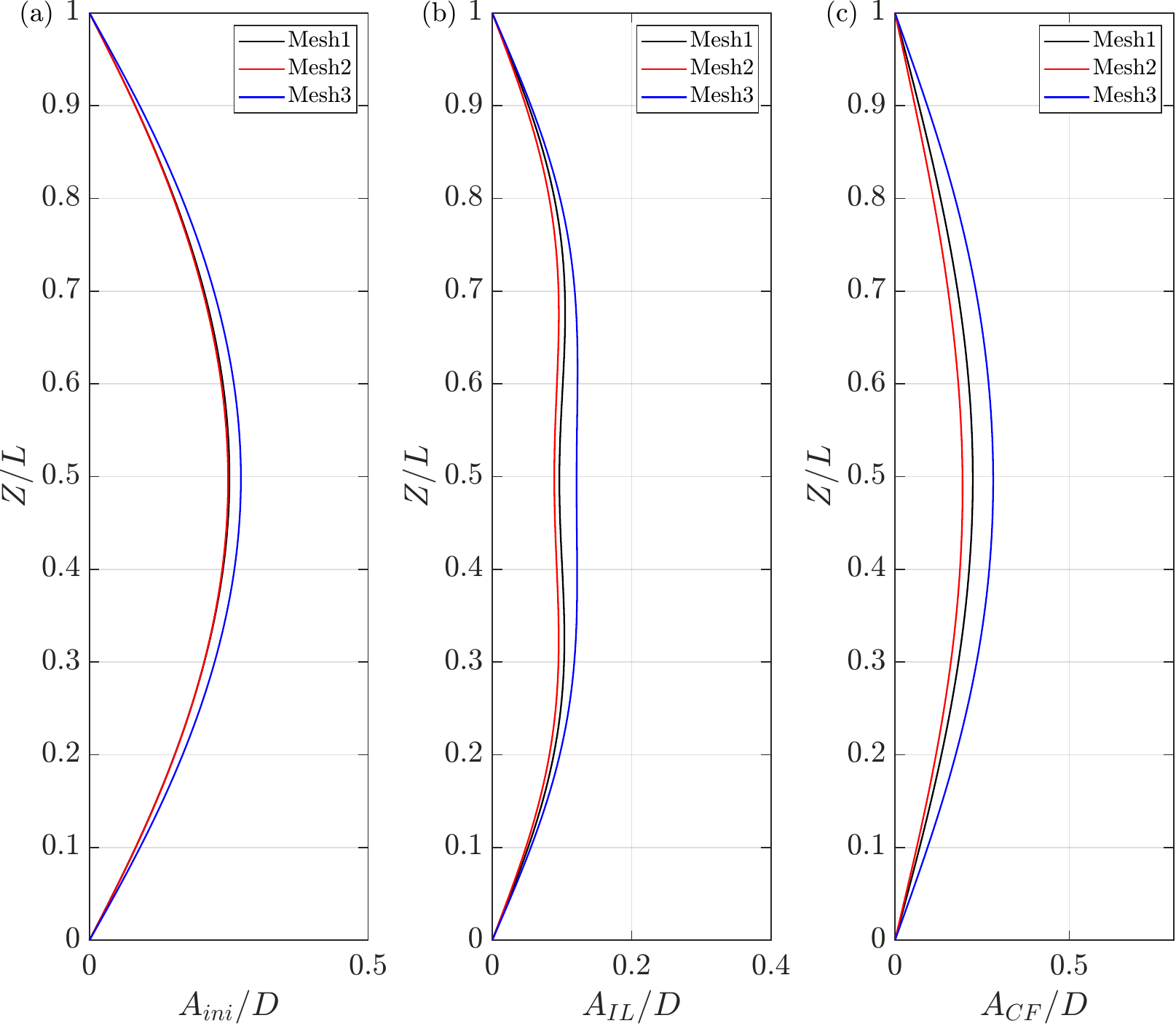}
    \caption{Independence test of the number of meshes used in the flow case of $U = \SI{0.40}{m/s}$: (a) initial IL displacement; (b) IL response; (c) CF response.}
  \label{meshnum}
\end{figure}

\cref{meshnum} show three different mesh number cases in uniform flow as examples. The cell numbers for mesh 1, mesh 2 and mesh 3 are 1418240, 1063040 and 909440, respectively. The results show that there is no difference among these cases, and mesh 2, as the median case, is chosen. We refer to the results of our previous numerical simulation studies to determine the appropriate mesh quality \citep{fu2022frequency,fu2023numerical}.

With respect to the number of strips, ensuring the distribution of three strips between nodes in the highest excitation mode can provide accurate prediction results \citep{chaplin2005blind}, and 20 strips are applied in the present study.

\section{Comparison of weak coupling and strong coupling results} \label{weakstrong}
The strong coupling algorithm is implemented for individual time steps to meet a high convergence standard, the FSI strong coupling algorithm with the Aitkenn relaxation method \citep{kuttler2008fixed} is shown in Algorithm \ref{alg:algorithmstrong}.

\begin{algorithm}[ht!]
	\caption{Strong coupling algorithm of FSI simulation with Aitken relaxation.}
	\label{alg:algorithmstrong}
    \KwIn{Solid field parameters; Flow field parameters.}
	\KwOut{Structural response $\boldsymbol{\xi}_s$ and flow velocity and pressure field $\boldsymbol{u}_f,p_f$.}  
	$t \leftarrow 0$\\
	$j_{in} \leftarrow 10,\; \epsilon_{in} \leftarrow 1\times 10^{-6}$
	
	\While{$t<t_{end}$}{
		$j \leftarrow 0,\;\gamma_j \leftarrow 0.5,\; \omega_j \leftarrow 1-\gamma_j$

	\While{$j<j_{in} \;||\; \epsilon>\epsilon_{in}$}{
	$\boldsymbol{F}_{f,j}^t$ $\leftarrow$ \textnormal{solve} $\mathcal{N}(\boldsymbol{u}_f,p_f)=0$
	
	 $\boldsymbol{\xi}_{s,j}^t$ $\leftarrow$ \textnormal{solve} $\mathcal{S}(\boldsymbol{u}_s)-\boldsymbol{F}_{f}=0 $
	
	\textnormal{Update mesh with} $\boldsymbol{\xi}_{s,j}^t$ \textnormal{at} $t$
	
	$\Delta \boldsymbol{\xi}_j \leftarrow \boldsymbol{\xi}_{s,j-1} - \boldsymbol{\xi}_{s,j}$
	
	$\epsilon \leftarrow |\Delta \boldsymbol{\xi}_j|$
	
	$\Delta \boldsymbol{\xi}_{j-1} \leftarrow \boldsymbol{\xi}_{s,j-2} - \boldsymbol{\xi}_{s,j-1}$
	
	$\gamma_j=\gamma_{j-1}+(\gamma_{j-1}-1)(\Delta\boldsymbol{\xi}_{j-1}-\Delta\boldsymbol{\xi}_j)\Delta\boldsymbol{\xi}_j/(\Delta\boldsymbol{\xi}_{j-1}-\Delta\boldsymbol{\xi}_j)^2$
	
	$\omega_j = 1 -\gamma_j$
	
	 $\boldsymbol{\xi}_{s,j}^t \leftarrow \omega_j\boldsymbol{\xi}_{s,j}^t  + (1-\omega_j)\boldsymbol{\xi}_{s,j-1}^t $
	 
	 $j  \leftarrow j+1 $
	}
		$t  \leftarrow t+\Delta t$}
\end{algorithm}

\cref{weakstrong} shows a comparison of the results of the weak and strong coupling algorithms. There is no significant difference between the two sets of results. However, three times as many simulations are required in the strong coupling case. \citet{causin2005added} analyzed the stability of FSI governing equations for flat-plate flows, yielding a characteristic equation $\chi(s)$ with:
\begin{equation}
    \chi(-\infty) = -\infty,\quad \chi(-1)=a+\frac{4}{\Delta t ^2}(\rho_f\mu_i-\rho_sh_s),
\end{equation}
where $a$ is a function of the material Young's modulus $f(E)$, $\rho_f$ and $\rho_s$ are the fluid and solid densities, $\mu_i$ is the eigenvalue of the added mass operator $\mathcal{M}_{A}$, and $h_s$ is the structural characteristic length. If $\chi(-1)>0$, the equation becomes unconditionally unstable. Thus, instability may occur when $\rho_s/\rho_f <1$ or with high Young's modulus values.

\begin{figure}[htb!]
    \centering
    \includegraphics[width=.5\textwidth]{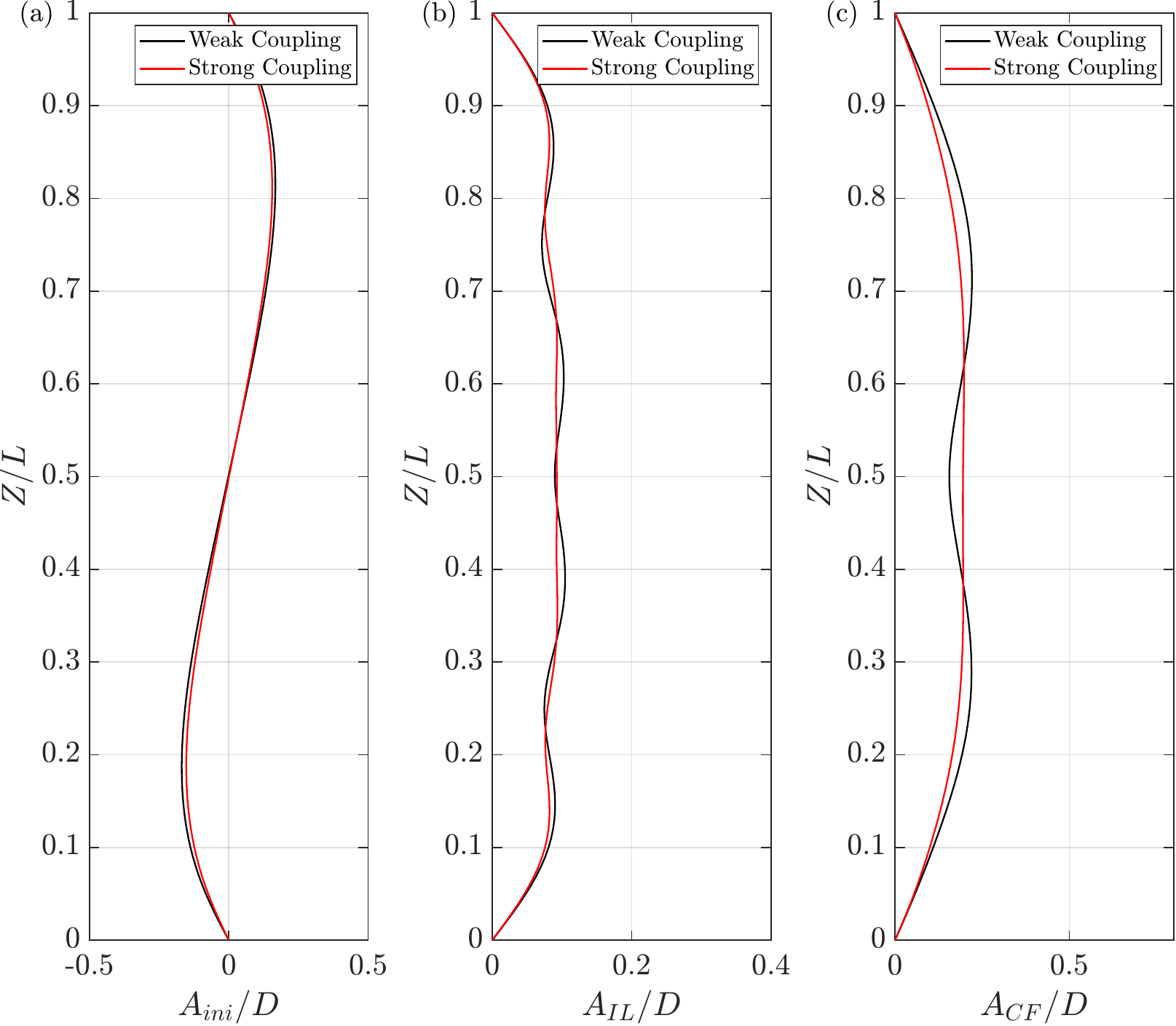}
    \caption{Comparison results of weak coupling and strong coupling algorithm.}
  \label{weakstrong}
\end{figure}

For VIV problems, flexible risers typically have mass ratios above $3.00$, and rigid cylinders have mass ratios above 1.00, therefore, the weak coupling algorithm is applicable for solving VIV problems. For practical flexible ocean engineering structures, in engineering software such as RIFLEX \citep{sintef2021riflex}, weak coupling is adopted for structural time-domain response simulations, which is considered to achieve sufficient engineering prediction accuracy. However, a submerged floating tunnel, a newly emerging circular structure in ocean engineering with a structural density similar to that of water, may require a strong coupling algorithm to perform accurate numerical simulations. This issue will be investigated in the future.

%\bibliography{mybibfile}

\end{document}